\documentclass[useAMS,usenatbib]{mn2e}
\usepackage{epsfig}
\usepackage{color}
\usepackage{subfigure}
\begin{document}
\def\Msun{\mbox{$M_\odot$}}
\def\Zsun{\mbox{$Z_{\odot}$}}
\def\lsim{\mathrel{\rlap{\lower3.5pt\hbox{\hskip0.5pt$\sim$}}
    \raise0.5pt\hbox{$<$}}}
\def\gsim{~\rlap{$>$}{\lower 1.0ex\hbox{$\sim$}}}

\def\Re{\mbox{$R_{\rm e}$}}
\def\RE{\mbox{$R_{\rm E}$}}
\def\mst{\mbox{$M_{\star}$}}
\def\SDM{\mbox{${\cal{S}}_{\rm DM}$}}
\def\St{\mbox{${\cal{S}}_{\rm tot}$}}
\def\gDM{\mbox{$g_{\rm DM}$}}

\newcommand{\goodgap}{\hspace{\subfigtopskip} \hspace{\subfigbottomskip}}

\def \tcb{\textcolor{blue}}
\def \tcr{\textcolor{red}}

\title[Correlations for GRB prompt and afterglow plateau
emissions]{Luminosity--time and luminosity--luminosity correlations for GRB
prompt and afterglow plateau emissions}

\author[M. Dainotti et al.]{M. Dainotti$^{1,2}$, V. Petrosian$^{1}$, R. Willingale$^{3}$, P. O'
Brien$^{3}$, M. Ostrowski$^{2}$, S. Nagataki$^{4}$\\
$^1$Astronomy Department, Stanford University, Via Pueblo Mall 382, Stanford,
CA, USA\\
$^2$Obserwatorium Astronomiczne, Uniwersytet Jagiello\'{n}ski, ul. Orla 171,
30-244 Krak\'{o}w, Poland\\
$^3$Department of Physics \& Astronomy, University of Leicester, Road Leicester
LE1 7RH, UK\\
$^4$Astrophysical Big Bang Laboratory, Riken, Wako, Saitama 351-0198, Japan\\}

\date{Accepted xxx, Received yyy, in original form zzz}

\maketitle

\begin{abstract}

We present an analysis of 123 Gamma-ray bursts (GRBs) with known redshifts
possessing an afterglow plateau phase. We reveal that $L_a-T^{*}_a$ correlation
between the X-ray luminosity $L_a$ at the end of the plateau phase and the
plateau duration, $T^*_a$, in the GRB rest frame has a power law slope
different, within more than 2 $\sigma$, from the slope of the prompt
$L_{f}-T^{*}_{f}$ correlation between the isotropic pulse peak luminosity,
$L_{f}$, and the pulse duration, $T^{*}_{f}$, from the time since the GRB
ejection. Analogously, we show differences between the prompt and plateau phases
in the energy-duration distributions with the afterglow emitted energy being on
average $10\%$ of the prompt emission. Moreover, the distribution of prompt
pulse versus afterglow spectral indexes do not show any correlation. 
In the further analysis we demonstrate that the $L_{peak}-L_a$ distribution,
where $L_{peak}$ is the peak luminosity from the start of the burst, is
characterized with a considerably higher Spearman correlation coefficient,
$\rho=0.79$, than the one involving the averaged prompt luminosity,
$L_{prompt}-L_a$, for the same GRB sample, yielding $\rho=0.60$. Since some of this correlation
could result from the redshift dependences of the luminosities,
namely from their cosmological evolution we use the Efron-Petrosian method to reveal
the intrinsic nature of this correlation. We find that a substantial part of the
correlation is intrinsic. We apply a partial correlation coefficient to the new de-evolved
 luminosities showing that the intrinsic correlation exists. 

\end{abstract}


\section{Introduction}
GRBs are the most distant and most luminous object observed in the Universe with
redshifts up to $z \approx 9.4$ and isotropic energies up to $10^{54}$ ergs.
Discovering  universal properties is crucial in understanding the processes
responsible for the GRB phenomenon. However, GRBs seem to be anything but
standard candles, with their energetics spanning over 8 orders of magnitude.
There have been numerous attempts to standardize GRB by finding 
some correlations among the observables, which can then be used for
cosmological studies.  Examples of these are the claimed correlations between
the isotropic total prompt emitted energy
$E_{iso}$ and the peak photon energy of the $\nu \times F_{\nu}$ spectrum $E_{peak}$.
\citep{Lloyd1999,Amati02,amati09}, the beaming corrected energy
$E_\gamma$  and $E_{peak}$ \cite{G04,Ghirlanda06}, the Luminosity $L$ and $E_{peak}$
\cite{S03,Yonekotu04}, and luminosity and variability $V$
\cite{FRR00,R01}.  However, because of the large dispersion in these
relations
\citep{Butler2007,Butler2009,Yu09} and possible impact of detector
thresholds, the utility of these correlation as a proxy for standard candle and
cosmological studies \citep{Shahmoradi09} have been
questioned \citep{Cabrera2007,Collazzi2008}.

In this paper we investigate whether some common features
may be identified in the light curves during both the prompt and afterglow
phases. 
A crucial breakthrough in this field has been the observation of GRBs by the Swift satellite, launched in 2004.
The on board instruments Burst Alert Telescope (BAT, 15-150 keV), 
X-Ray Telescope (XRT, 0.3-10 keV), and  Ultra-Violet/Optical Telescope (UVOT,
170-650 nm), provide a broad wavelength coverage and a rapid follow-up of the
afterglows.
Swift has revealed a complex behavior of the light curves \citep{OBrien06,Sak07}, where one can distinguish two, three or even more
segments in the afterglow. The second segment, when it is flat, is called the
plateau emission. Investigating the X-ray afterglow  \cite{Dainotti2008, Dainotti2010} 
discovered a power-law anti-correlation between the rest frame  time $T^*_a$,
when the plateau ends and a power-law decay phase begins, and $L_a$,  the
isotropic X-ray luminosity at $T^*_a$.%
\footnote{Here,
and subsequently, $*$ denotes the rest frame quantities. These quantities are
obtained
by fitting
the light curves to the 
phenomenological Willingale et al. (2007) model, hereafter called W07, and all
luminosities and
the respective derived energies are for an assumed isotropic emission. To
simplify the
notation  we  omit the subscript `iso".} 
This correlation has also been
reproduced independently by other authors with slopes within 1 $\sigma$ of the
above value.
\citep{Ghisellini2009,Sultana2012} 
\footnote{A luminosity-time correlation has been found
also for short GRBs with extended emission \citep{Dainotti2010} and future
perspective will be the investigation of this class of GRBs within the model of
Barkov \& Pozanenko (2011)}.
However, some of this correlation is induced by the
redshift dependences of the variables. More recently, Dainotti et al. (2013a)
have demonstrated that after correcting for this observational bias there
remains a significant (at $12$ sigma level) anti-correlation with the intrinsic slope
$b=-1.07_{-0.14}^{+0 09}$.

The $L_a-T^{*}_a$ anti-correlation has been a useful test
for theoretical interpretation of GRB models involving accretion
\citep{Cannizzo09,Cannizzo11}, a magnetar
\citep{Dall'Osso,Bernardini2012a,Bernardini2012b,Rowlinson2010,Rowlinson2013,
Rowlinson2014}, the long-lived reverse shock models \citep{Leventis2014,Erten2014a}, 
and other additional models such as the prior emission model
\citep{Yamazaki09}, the unified GRB and AGN model \citep{Nemmen2012} and the
induced gravitational collapse scenario \citep{Izzo2012}. There are several
models, e.g the photosperic emission model \citep{Ito2014}, that can account for
this observed correlation. In addition, Dainotti et al. (2011a) attempted to use this relation as a
redshift estimator and  \cite{Cardone09,Cardone2010,Postnikov2014}, have
used it for cosmological studies. But \cite{Dainotti2013b} have described some 
caveats on the use of non-intrinsic correlations to constrain cosmological
parameters. Dainotti et al. (2015) used this correlation to evaluate the redshift-dependent ratio $\Psi(z) = (1 + z)^{\alpha}$ of the
GRB rate to the star formation rate.

The aim of this paper is to compare similar luminosity-duration correlations in the light curve of the prompt emission with the afterglow ones. This may shed light on the
relative energizing, dissipation and radiative processes of afterglow and
prompt emission.
Dainotti et al. (2011b) have
demonstrated the existence of a tight correlation between the afterglow
luminosity $L_a$ and the average $L_{prompt}$
luminosity over all the prompt emission phase.
 Moreover, Qi (2010) has
discovered for the first time the existence of luminosity duration anti-correlation in the prompt emission.
Later, Sultana et al. (2012) used a sample of 12 GRBs to show that the burst
peak isotropic luminosity, $L_{peak}$, and the spectral lag, $\tau$,
distribution continuously extrapolates into the $L_a-T^*_a$ distribution, with a
common correlation slope close to $-1.0$. The authors conclude that, if indeed
the underlying physics is common, it should be of kinematic origin. 
Because the lag time $\tau$ is somewhat different variable than the
durations in the light curves, we propose a more direct
comparison between the $L_{a}-T^*_{a}$ correlation and the $L_{f}$-$T^{*}_{f}$
where $L_{f}$ and $T^{*}_{f}$ stand for the peak luminosity and pulse
width of individual gamma ray pulses in the prompt emission. We here use the same notation of $L_f$ and $T_f$ following the original notation of Willingale et al. (2010). Because the W07 model masks out the flares in the light curve, we use the Willingale et al. (2010) model (hereafter W10) which is
more appropriate for dealing with individual pulses.
In the next section we present the theoretical motivations for this data
analysis and what can be learned from the results. In \S 3 we describe the
modeling of the light curves ans in \S 4 we describe the data
analysis. The results on the luminosity duration correlation are presented
in  \S 5 and a brief summary and discussion is presented in \S 6.

\section{Theoretical Motivation}

To start we summarize some selected models in the
literature which address the luminosity-duration correlations and
attempt to explain the observed luminosity prompt-afterglow correlations.
 
1) The commonly invoked cause of the plateau formation by continuous energy
injection into the GRB generated forward shock leads to an efficiency crisis for
the prompt mechanism as soon as the plateau duration exceeds $10^{3}$ seconds.
Hascoet et al. (2014) studied two possible alternatives: the first one within the framework of the standard forward shock model but
allows for a variation of the microphysics parameters to reduce the radiative efficiency at
early times; in the second scenario the early afterglow results from a long-lived reverse shock in the forward shock scenario. In both scenarios
the plateaus following the prompt-afterglow correlations can be obtained under the condition
that additional parameters are added. In the forward shock scenario the preferred
model supposes a wind external medium and a microphysics parameter $\epsilon_e$,
the fraction of the internal energy that goes into electrons (or positrons) and can in
principle be radiated away. This varies as $n^{−\nu}$ (where $n$ is the external density), with $\nu
\approx 1$ to obtain a flat plateau. They conclude
that acting on one single parameter can lead to the formation of a plateau that
also satisfies the observed prompt-afterglow correlations presented in Dainotti
et al. (2011b). Another possibility presented by Hascoet et al. (2014) is the
reverse shock scenario, in which the typical Lorentz factor of the ejecta should
increase with burst energy to satisfy the prompt-afterglow relations, more in particular the ejecta must
contain a tail of low Lorentz factor with a peak of energy deposition at $\Gamma \ge 10$. \newline

2) Van Eerten (2014b) shows that the observed $L_{prompt}-L_{afterglow}$ correlations rule out basic thin
shell models but not basic thick ones. In the thick shell case, both forward
shock and reverse shock outflows are shown to be consistent with the correlations,
through randomly generated samples of thick shell model afterglows. A more
strict approach with the standard assumption on relativistic blast waves  is used in the
contexts of both thick and thin shell models. In the thin shell model, the
afterglow plateau phase is the result of the pre-deceleration emission from a
slower component in a two-component or jet type model. For thick shells, the
plateau phase results from energy injection either in the form of late central
source activity or via additional kinetic energy transfer from slower ejecta
which catches up with the blast wave. It is shown that thin shell models can not
be reconciled with the observed LT correlation and, then, it is inferred the
existence of a correlation between the plateau end time and the ejecta
energy that is not seen in the observational data. However, this does not mean
that acceptable fits using a thin shell model are not possible, it might even be
possible to successfully fit all the bursts with plateau stages. Thick shell
models, on the other hand, can easily reproduce the LT correlation even if
uncorrelated values for the model parameters are applied in modeling. In this
context it is difficult to distinguish between forward shock and reverse shock
emission dominated models, or homogeneous and stellar wind-type environments.\newline

3) A supercritical pile-up model \citep{Sultana2013} provides an explanation for
both the steep-decline-and-plateau or the steep-decline-and-power-law-decay structures of the GRB afterglow phase, as observed in a large number of
light curves, and to the LT relation. Since in this model,
 the detailed calculations an estimate of the Energy of the prompt is needed, it would be relevant to evaluate if the $L_{prompt}-L_{afterglow}$ and the $L_{peak}-T_{peak}$
relations, as defined here, can be reproduced.\newline

4) Ruffini et al. (2014) show that the induced gravitational collapse paradigm
is able to reproduce the $L_a-L_{prompt}$ relations very tightly. More in general, this model
addresses the very energetic ($10^{52}–-10^{54}$ erg) long GRBs associated with
Supernovae. They manage to reproduce the lightcurves giving different scenarios for the circumburst medium, 
with either a radial structure for the wind \citep{Guida2008} or with a fragmentation of the shell \citep{Dainotti2007} thus well fitting the afterglow plateau and the prompt emission. \newline

Given this wide  possible theoretical interpretations it is important
to take into consideration additional information from the observational
correlations presented in this paper. This can help to provide new constraints for the physical models of GRB
explosion mechanism.\newline

\section{Modeling the GRB light curves}\label{Willingales}

\begin{figure}
\includegraphics[width=1.03\hsize,height=0.45\textwidth,angle=0,clip]
{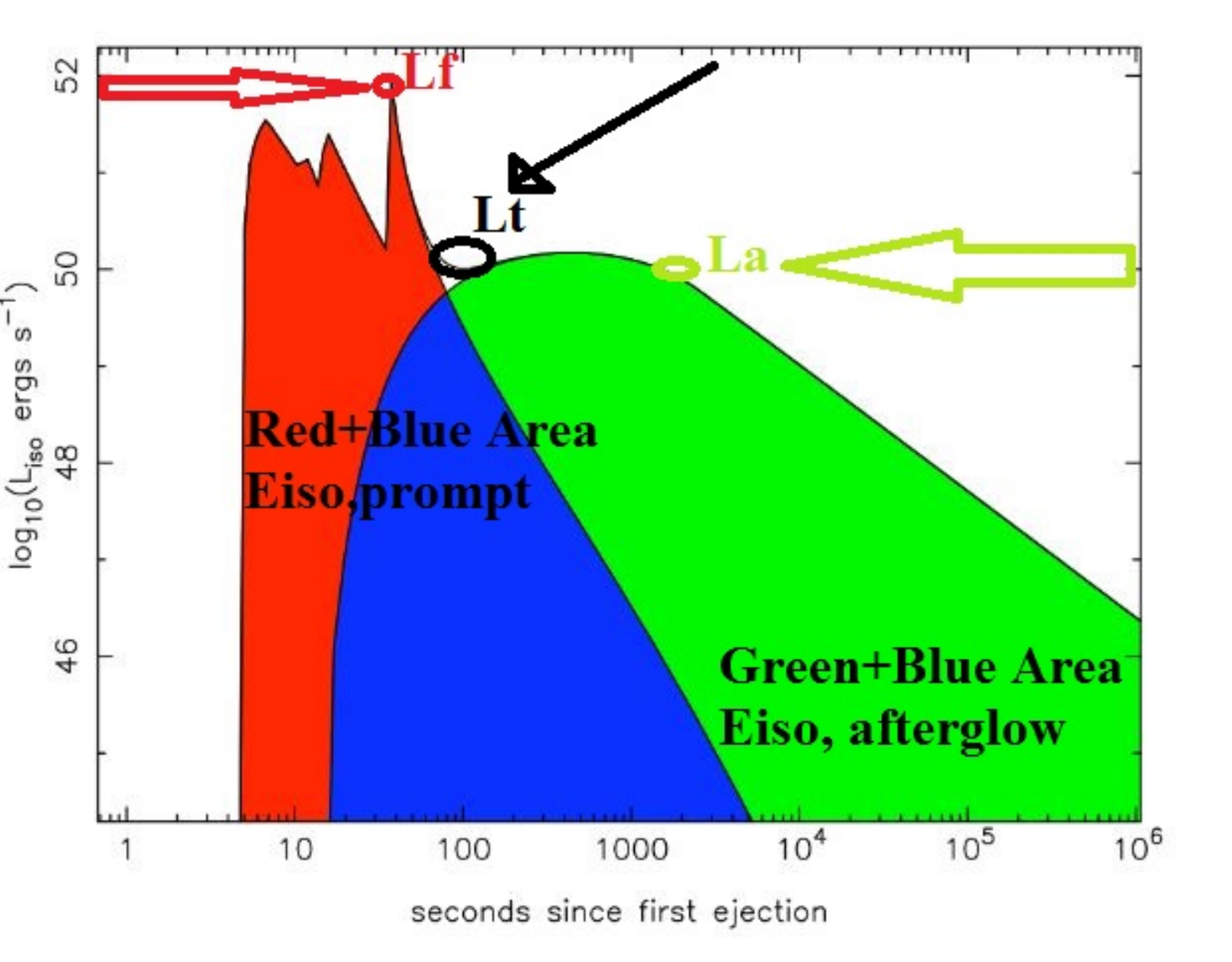}
\caption{A schematic light curve which illustrates how the prompt and afterglow
emission components are integrated to obtain the respective energies within the
W010 model. The red + blue area is proportional to the energy of the prompt
emission, where we also indicated the time $T_{f}$, the duration of the pulse
since the time of the GRB ejection. The green one + the blue area indicates the
afterglow's energy, where $T_a$ is the time of the end of the plateau emission.
In the joint area (blue) $T_t$ is the time where the luminosities of the
decaying prompt emission and the afterglow emission are equal. The solid line is the total
luminosity. 
\label{fig1}} 
\end{figure}

Usually the X-ray light curves of afterglows observed by XRT are modeled
using a series of power laws segments plus pulses; see
e.g. \citep{Evans2009,Evans2010,Evans2014,Margutti2013}. Here we use a different
approach whereby we fit the light curves to the analytic functional
forms of W10, which, as mentioned above, is an improved version of W07 and  fits
the complete BAT+XRT light curves without masking the X-ray flares. This
procedure uses somewhat physically motivated pulse profile for the prompt
emission, based on the spherical expanding shell model
\citep{Ryde2002,Dermer2007}, where the shells are energized during the rise of
the pulse and the decay phase of the pulse involves emission generated further away from the line of sight that arrive latter and with a
smaller Doppler boost.
 
The peak luminosity and pulse width of the individual pulse are denoted 
as $L_{f}$ and $T_f$ while $L_a$ and $T_a$ refer to the afterglow values define
above. Fig. \ref{fig1} shows these quantities for a schematic light curve. We
also determine the total energy fluence $E$ for pulses and the afterglow
phase. The rest frame  times $T^{*}_{f}$ and $T^{*}_a$ 
represent the times when the respective energy supply is switched off.

\subsection{Nomenclature} \label{nomenclature}

For clarity we report a summary of the nomenclature adopted in the paper (c.f.
Fig. \ref{fig1}). All times described below are given in the observer frame,
while with the upper index $^*$ we denote in the text the observables in the
GRB rest frame. All considered energies and luminosities are derived assuming
the isotropic emission.
\begin{itemize}
\item{$T_{peak}$, is the peak luminosity time in the prompt emission, measured
since the start of the burst. Its corresponding luminosity is $L_{peak}$.}
\item{$T_f$ is the pulse peak time in the prompt emission computed from the GRB
ejection time, $T_{ej}$. Its corresponding luminosity is $L_{f}$.}
\item{$T_{prompt}$ is the sum of all the pulse peak times, $T_f$, for each GRB in the prompt} 
\item{$T_{90}$ is the time between the $5\%$ and $95\%$ of the energy released
in the GRB prompt phase.}
\item{$T_{45}$ is the time between the $5\%$ and $50\%$ of the energy released
in the GRB prompt phase.}
\item{$L$ and $T$ indicate the luminosity and time which can be either for the
prompt ($L_f$ or $L_{peak}$; $T_f$ or $T_{peak}$) or the afterglow ($L_a$;
$T_a$) emission.  The equivalent energy-duration $E$ and $T$ relations are also
considered.}
\item{$E_{min}$ and $E_{max}$ are respectively the minimum and maximum energy
in the band pass of the instrument. For the XRT a respective range is  ($0.3$,
$10$) keV, while for the BAT  it is ($15$, $150$) keV.}
\end{itemize}

\section{Data analysis} \label{data analysis}

\begin{figure}
\includegraphics[width=0.99\hsize,height=0.30\textwidth,angle=0,clip]{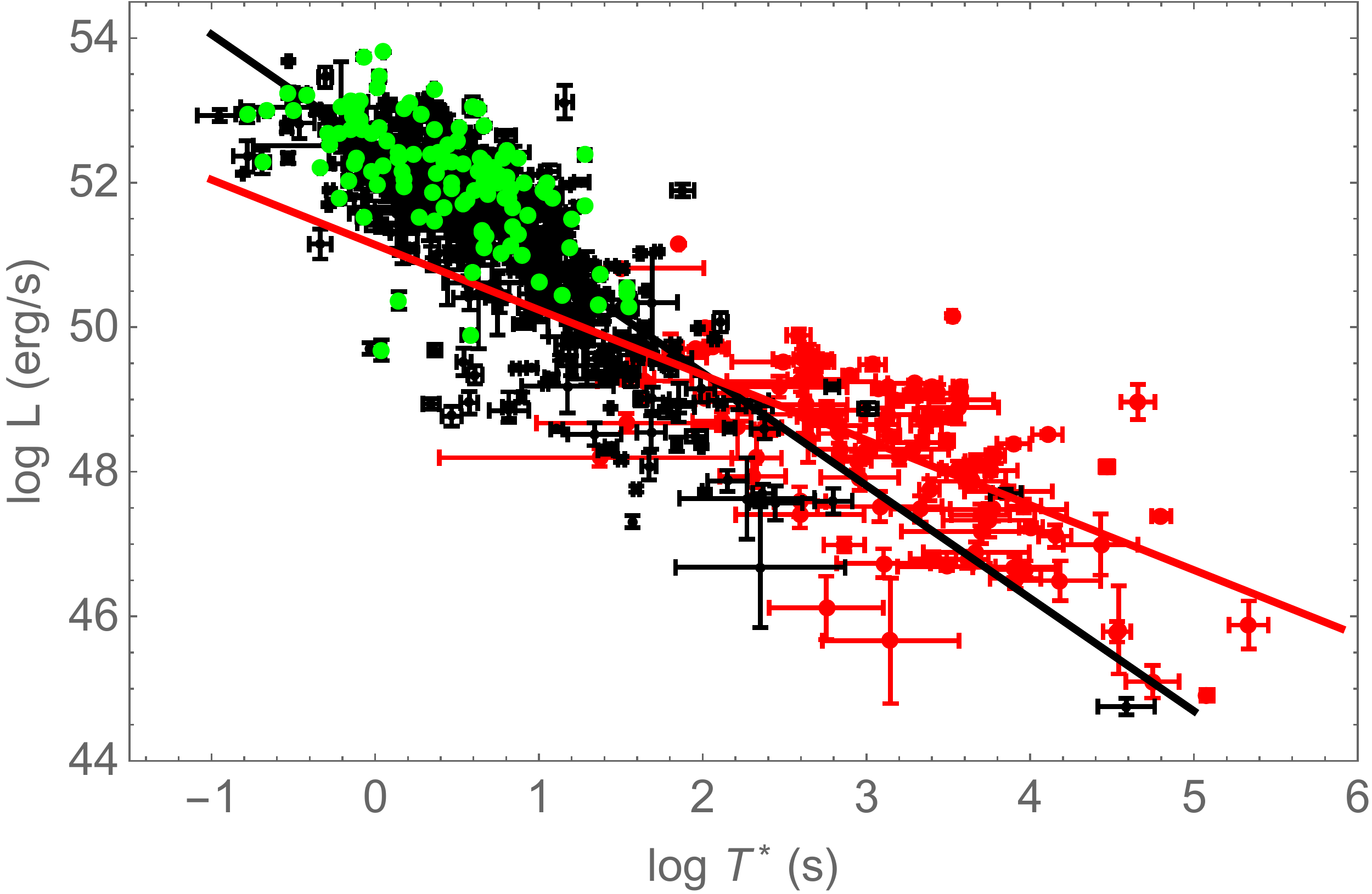}
\includegraphics[width=0.99\hsize,height=0.30\textwidth,angle=0,clip]{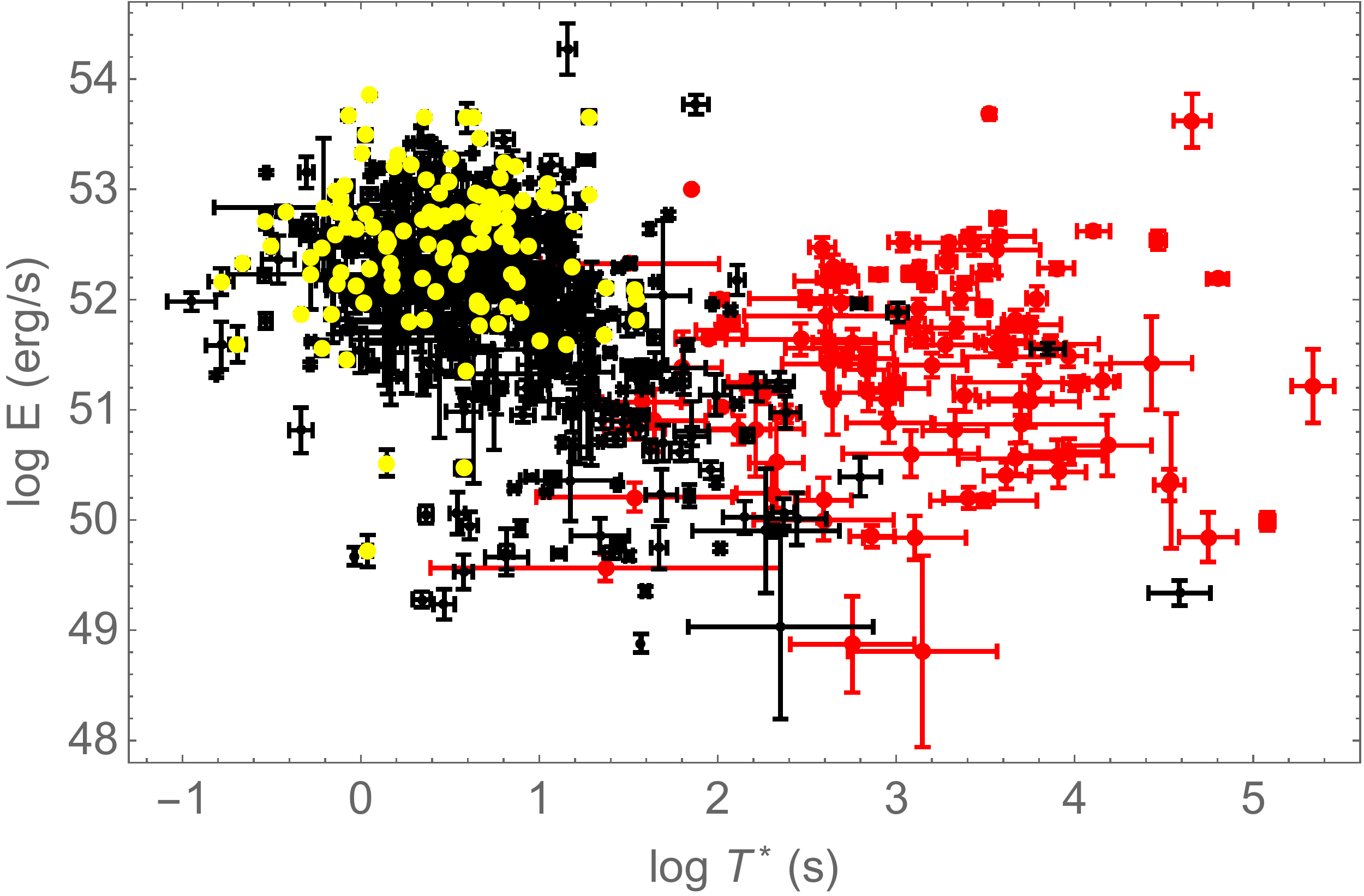}
\includegraphics[width=0.99\hsize,height=0.30\textwidth,angle=0,clip]{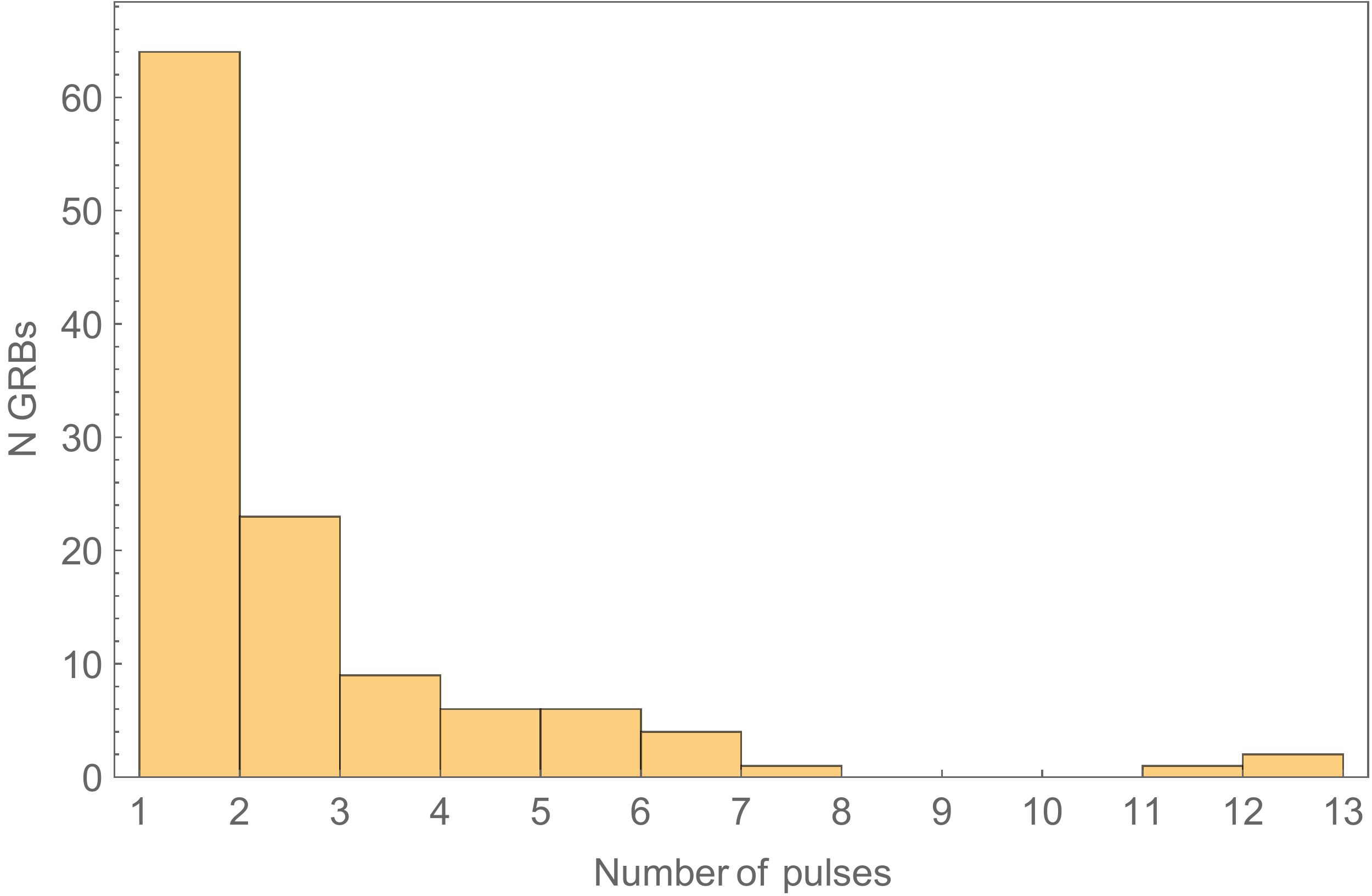}
\caption{Distributions of $L$ vs $T^*$  (upper panel) and $E$ vs. $T^*$ (middle
panel) for each single pulse both in the prompt (black symbols) and in the
afterglow (red symbols) emissions. $L$ and $E$ are  equal to $L_{f}$ and $E_{f}$
for the prompt emission pulses, while being equal to $L_a$ and
$E_{afterglow}=L_a*T^{*}_a$ for the afterglows, and, respectively, the time
$T^*$ represents $T^*_{f}$ for the prompt emission pulses and $T^*_a$ for the
afterglow phase. The green points represent the highest luminosity prompt
emission pulses ($T_{Lmax}$,$L_{max}$), while the yellow ones represent
($T_{Emax},E_{max}$).  In the bottom panel, we show a distribution of the number
of maximum luminosity pulses in the GRB pulse histogram. 
\label{fig2}}
\end{figure}

We have analyzed the sample of long GRBs with known redshifts detected
by Swift from January 2005 up to September 2011, for which the light curves
include early XRT data. The redshifts $z$ are taken from J. Greiner's Web
site \footnote{http://www.mpe.mpg.de/~jcg/grbgen.html} and from Xiao \& Schaefer
(2009). Among these GRBs we have selected 123 with early XRT coverage for the
fitting. Thus, the BAT-XRT combined data give us almost continuous monitoring of
the GRB varying emission. On the other hand, we rejected all bursts where a gap
in the XRT coverage reveal flares with only partial coverage, missing the turn
on, the peak and/or the decay phases. For both prompt and afterglow components we compute the luminosity 
in the appropriate energy bandpass, $(E_{min}, E_{max})$, as:

\begin{equation}
L(E_{min},E_{max},t)= 4 \pi D_L^2(z) \, F (t) \cdot K(E_{min},E_{max}),
\label{eq: lx}
\end{equation}

\noindent
where $D_L(z)$ is the luminosity distance computed in the flat $\Lambda$CDM
cosmological model with $\Omega_M = 0.291$ and $h=0.70$ in units of $100$ $km$
$s^{-1}$ $Mpc^{-1}$, $F$ is the measured X-ray energy flux and \textit{K} is the
\textit{K}-correction for the cosmic expansion \cite{B01}:

\begin{equation}
K=\frac{\int_{E_{min}/(1 + z)}^{E_{max}/(1 + z)}{\Phi (E)
dE}}{\int_{E_{min}}^{E_{max}}{\Phi (E) dE}},
\label{eq: kcorrection}
\end{equation}

\noindent

where the energy spectrum $\Phi (E)$ of the afterglows is described
by a simple power law $\Phi (E) = E^{-\beta_a}$, while the one of the prompt pulses
by the Band function \citep{band93}.
\footnote{For the prompt pulses $\beta_{pulse}$ is the low
energy index of the Band spectrum and the spectral fits are calculated separately from
the afterglow ones within the ($E_{min}$, $E_{max}$) = ($15$-$150$) keV in the
$4$ BAT energy channels ($15-25$ keV, $25-50$ keV, $50-100$ keV, $100-150$ keV).
We point out here that the spectrum is not extrapolated at low
energy in the afterglow, but it has been computed separately. Moreover, in the afterglow phase generally there is no spectral evolution; 
few bursts which show spectral evolution are not in our list of GRBs.}

We also employ another way to compute $L_{peak}$, instead of using the functional form of Willingale et al. (2010), 
we follow Schaefer et al. (2007) and Eq. \ref{eq: lx}, using the
brightest peak flux over $1$ sec interval \footnote{In our sample there is always
a peak flux defined for $1$ sec interval.}. For the functional form for the
spectrum, we use either a power-law (PL) or a power law with a cutoff (CPL),
depending on the best $\chi^{2}$ fit presented in the Second BAT Catalog
(differently from the approach used in W010 in which the Band function for the
pulse profile is adopted). All of the BAT spectra are acceptably fitted by
either a PL or a CPL model. The same criterion as in the first BAT catalog,
$\Delta\chi^{2}$ between a PL and a CPL fit greater than 6 ($\Delta\chi^{2}
\equiv \Delta\chi_{PL}^{2}-\Delta\chi_{CPL}^{2}$), was used to determine if the
CPL model is a better spectral model for the data. Note that none of the BAT
spectra show a significant improvement in $\Delta\chi^{2}$ with a Band function
\citep{band93} fit compared to that of a CPL model fit. For GRBs not presented
in the Catalog we have chosen the spectral energy distribution as a function
that gives the best $\chi^{2}$ according to the Swift Burst Analyzer,
$http://www.swift.ac.uk/burstanalyser/$ \citep{Evans2009}, which are consistent
with the approach of the second BAT catalog. 
{\bf For the derivation of the pulse energy we integrated the fitted model
luminosity curve for each pulse as follows:

\begin{equation}
E_{pulse}= \int_{T_0}^{T_{end}} {4 \pi D_L^2(z) \, F (t) \cdot K(E_{min},E_{max}) dt},
\label{eq: energy}
\end{equation}

where $T_0=T_f-T_{ej}$ following the W010 notation, while $T_{end}$ is the time end of the pulse width, for these definitions see section \ref{nomenclature}}. The energy is presented on the lower panel of Fig. \ref{fig2}.

In what follows we use the above data for comparing the prompt and afterglow
characteristics and correlations. 

\section{Results}
\label{results}

The results are presented in  Fig. \ref{fig2}. The top panel shows the
luminosity-time, LT, scatter diagram including both pulses ($L_f-T^*_f$, black points) and the
afterglow ($L_a-T^*_a$, red points) while the middle panel shows the energy, ET, scatter
diagram, where the afterglow energy is calculate as $E_a=L_a*T^{*}_a$. The
lower panel shows the distribution on number of pulses per GRB. For each GRB
we also show the brightest luminosity (integrated over 1 s) $L_{f,max}$ (green) and
$E_{peak,max}$ (yellow) taken as the maximum $L_f$ and $E_{peak}$ among the pulses of a given GRB. 
\footnote{We note that the catalog uses a power law or a power law with an
exponential break, instead of the Band function, for the spectral fitting.}
We first note that using the new and larger sample we have repeated the
analysis carried out in Dainotti et al. (2013a) on the $L_a-T^{*}_a$ 
correlation and find similar results. A fit to this
relation $\log L_a = \log a + b \cdot \log T^*_a$ using a Bayesian method
\citep{Dago05} yields the observed intercept $\log a_{\rm plateau}=51.14 \pm 0.58$ and slope $b_{\rm plateau}=-0.90_{-0.17}^{+0.19}$ and the probability of the correlation occurring by chance for an uncorrelated sample is $P \approx 10^{-35}$ \citep{Bevington}. 

\subsection{The $LT$ Correlations}

As shown in the upper panel of Fig. \ref{fig2}) there is a strong
$L-T^{*}$ anti-correlation for both the prompt pulses and the plateau.
Linear fits to  $\log L$ vs $\log T$ using the
D'Agostini method \citep{Dago05} described in the Appendix, yields slopes and
intercepts respectively to be $b_{\rm prompt}=-1.52_{-0.11}^{+0.13}, \log a_{\rm prompt}= 52.98 \pm
0.08$ erg/s for the prompt pulses, and $b_{\rm
plateau}=-0.90_{-0.17}^{+0.19}, \log a_{\rm plateau}=51.14 \pm 0.58$ for the
plateau. The slopes differ almost by $3 \sigma$ implying a significance
difference at least in the observed correlations. 
 More credence can be given to this results, because we have used the same
W10 method for determining the luminosities and duration for both prompt and afterglow 
components.
This makes the comparison between $L_{f}$-$T^{*}_{f}$ and $L_{a}-T^*_{a}$ well defined. It has
already been demonstrated within the context of W07 that both prompt and
afterglow emission can be represented by the same functional form.  The
underlying hypothesis, which we test here, is that the plateau can be
considered as a single flare with origin similar to the peaks of the prompt
emission.
Another way to look at this correlation is to consider the energy-duration
correlation, where the energy is computed integrating the pulse shape over the
pulse width. As expected we see much shallower relation for energies than
luminosities. The prompt pulses show still a
weak anti-correlation, but there is no correlation between 
$E_{a}$ and $T^{*}_a$ for the plateau.
The prompt emission pulses and the plateau data occupy two distinctive regions on the
energy-duration plane. The pulses are short and have slightly higher average
energy as compared to the plateau, which are in average $214$ times longer.
However, there is continuity in the distribution between prompt and plateau
pulses, namely there is also a small region of overlapping among the two phases.

For clarity, in the lower panel of Fig. \ref{fig2}, we present the distribution of
$L_{max}$, which is the maximum value of $L_{peak}$ in a burst, in
correspondence of its peak number, namely at which the peak occurs. We note that
the majority of $L_{max}$ occur between the first and second peaks of the prompt
emission, only in rare cases $L_{max}$ correspond to a peak number which exceeds
$10$. 

\subsection{Spectral Features of the pulses}
\label{spectral features}

\begin{figure}
\includegraphics[width=0.49\hsize,height=0.23\textwidth,angle=0,clip]{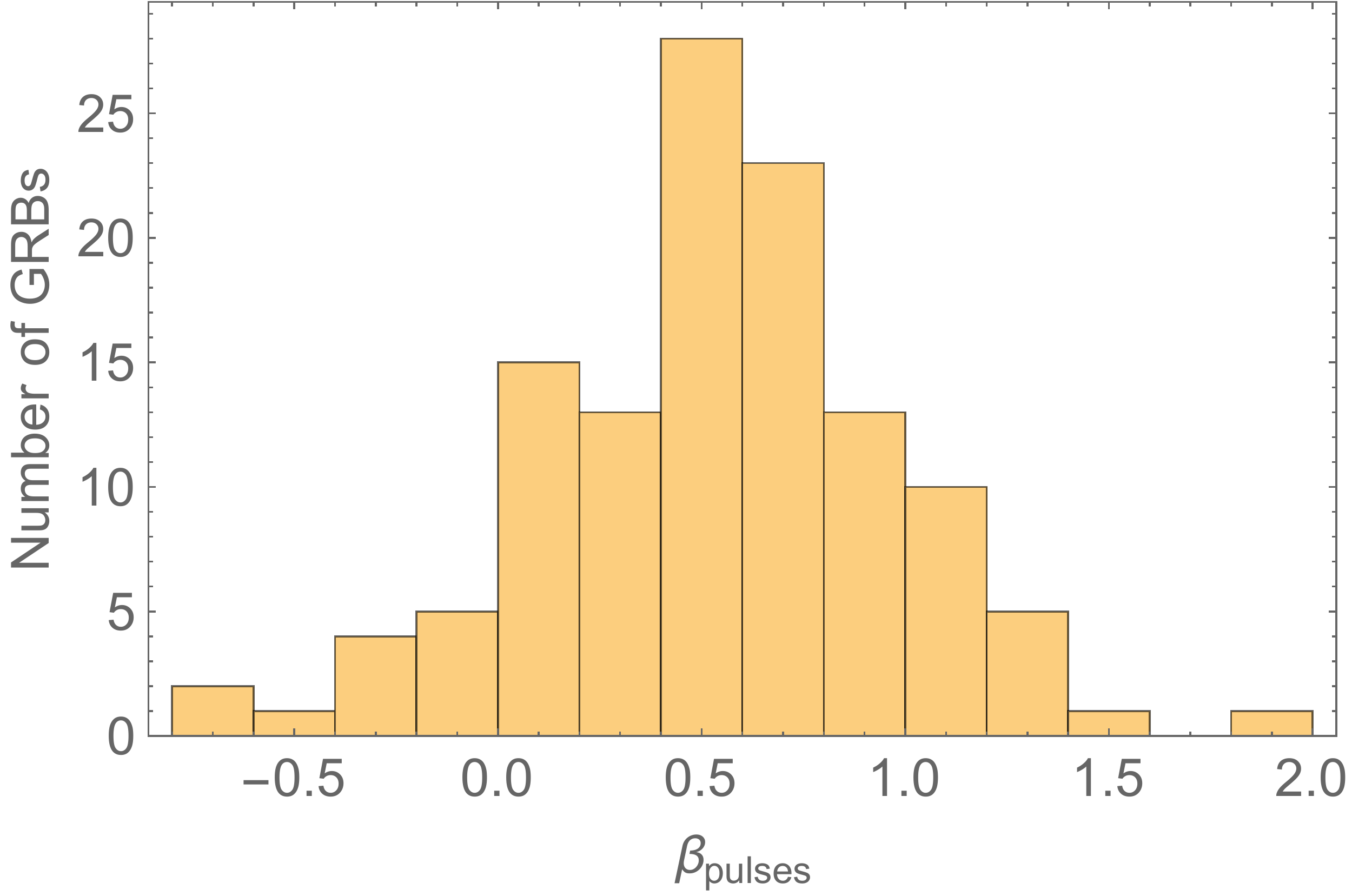}
\includegraphics[width=0.49\hsize,height=0.23\textwidth,angle=0,clip]{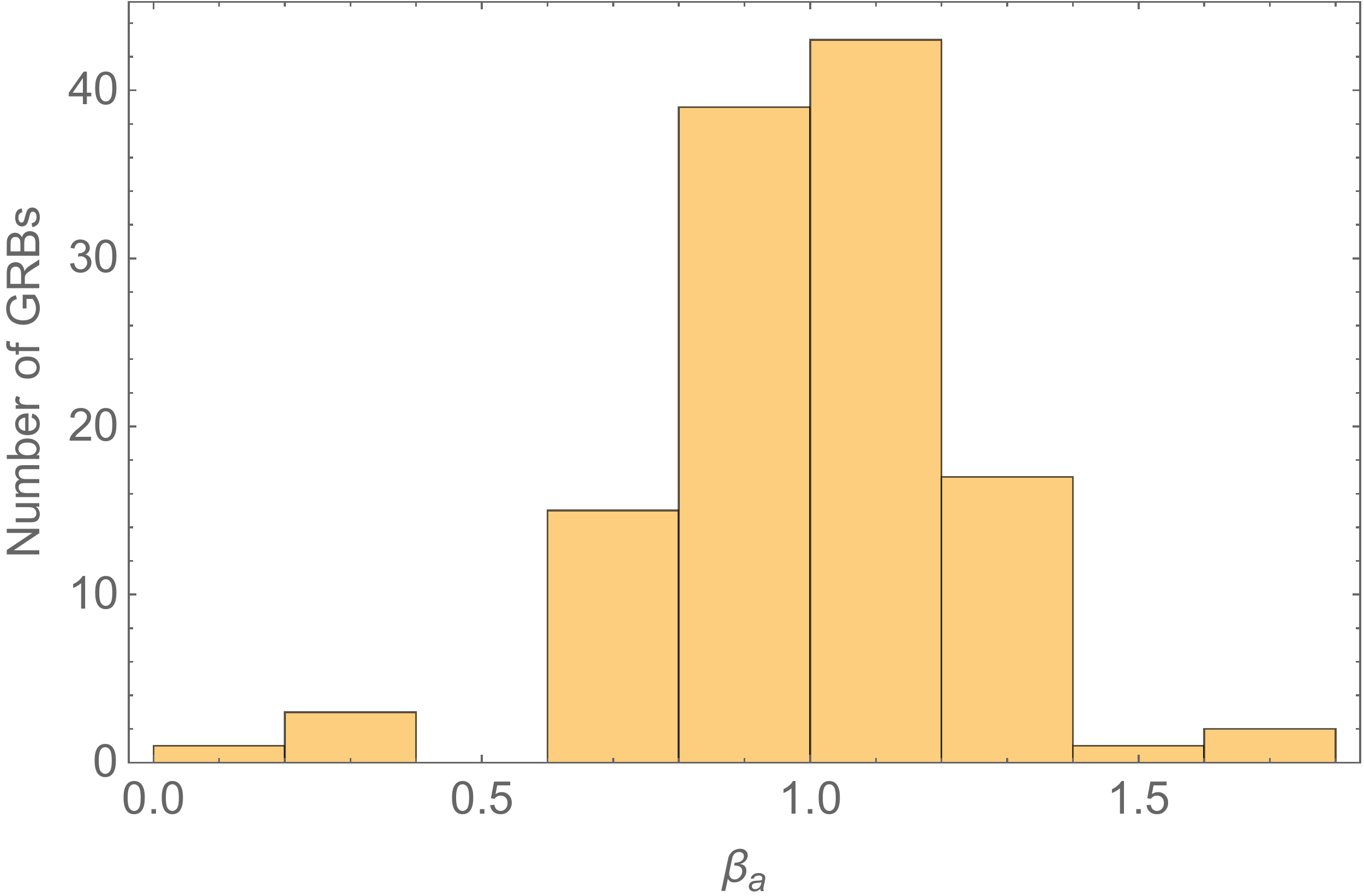}
\caption{Spectral index distributions for the prompt emission pulses,
$\beta_{pulses}$ (left panel); the pulses in the afterglow phase (right panel),
$\beta_a$. We represent all the pulses both in the prompt and in the afterglow
emission.}
\label{fig3} 
\end{figure}

\begin{figure}
\includegraphics[width=0.90\hsize,angle=0,clip]{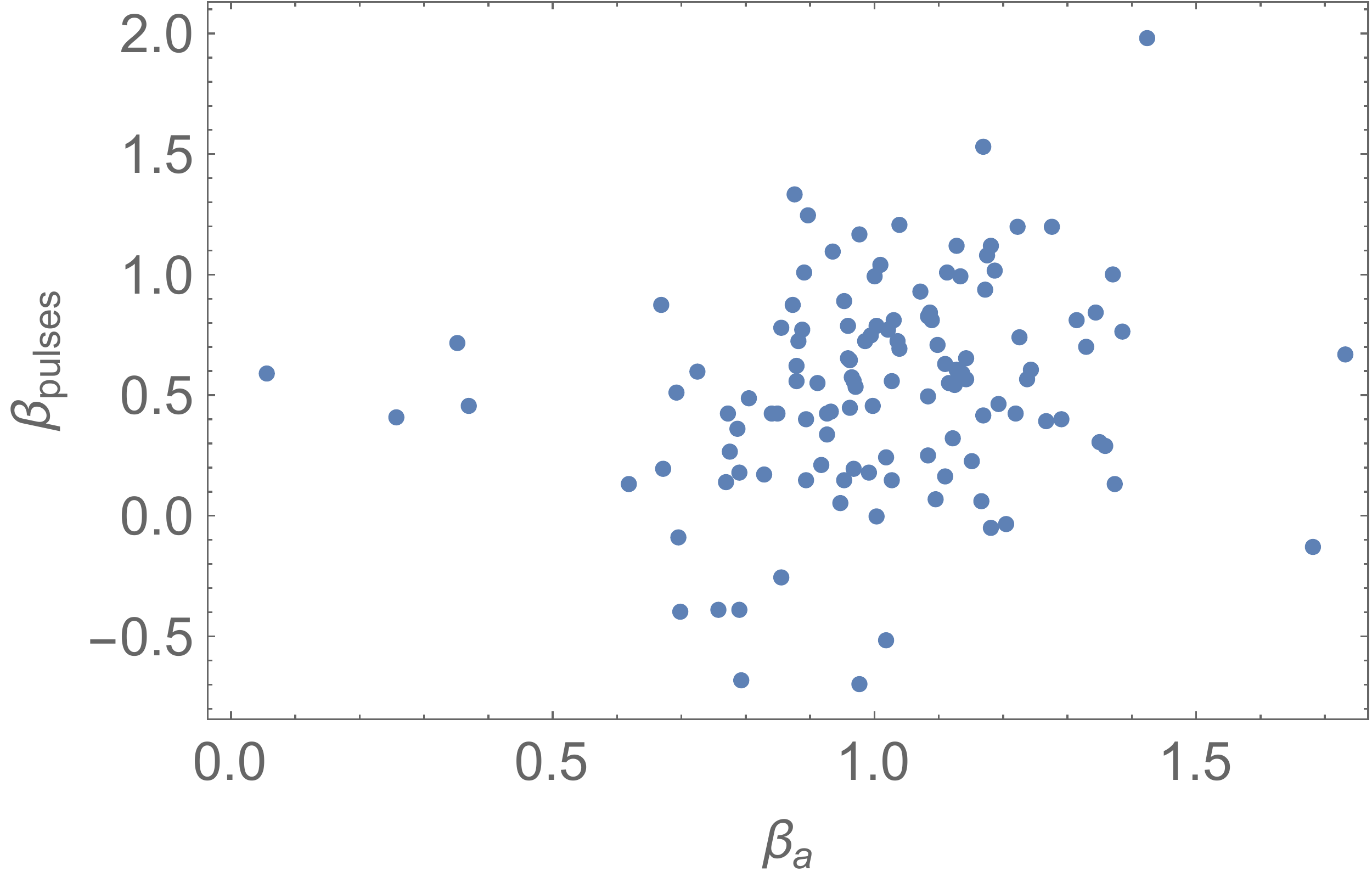}
 \caption{Spectral index distribution of the averaged $\beta_{pulses}$ among the
pulses in each GRB vs $\beta_a$ both computed within the W010 model. We note
that there is no correlation among the two distributions.}
\label{fig4} 
\end{figure}

We now compare the spectral characteristics. Fig. \ref{fig3} shows the
distribution of spectral indexes of 628 prompt pulses and 123 from the afterglows.
The two distributions are significantly different. The distribution of the prompt pulse
indexes is broader than that of the afterglow. As mentioned above, the spectral
index $\beta_{a}$ does usually not evolve
\citep{Evans2014}, it is constant over the plateau phase and later during the
afterglow decay phase, while the values of $\beta_{pulses}$ may vary during the
prompt emission phase. On Fig.\ref{fig4} we plot the average index of
prompt pulses in each source versus the afterglow index. There seem to be very
little correlation between the two indexes with most GRBs having a harder
prompt than afterglow spectra.
 
Moreover, the spectral parameters do not correlate strongly with the
other parameters we have introduced so far such as $E$, $L$ and the various
timescales.  
When inspecting the Fig. \ref{fig3}, the spectral index of the pulses evolves and this evolution has
been considered in the pulse model fit. Here, the spectrum of each single
pulse has been computed. We note that the $\beta_{pulses}$ computed for each
pulse have wider distributions than the typical values, integrated over
$T_{90}$, of $\beta$ in the prompt phase. These differences in spectral index do
not imply necessarily or justify a difference in the luminosity-time correlation
slopes. In fact, spectral breaks and spectral evolution can in principle explain
their diverse distributions.

\subsection{Luminosity-Luminosity Correlation}\label{luminosity-luminosity
correlations}

\indent We now compare prompt energy- afterglow energy and prompt
luminosity- afterglow luminosity correlations.

\begin{figure}
\includegraphics[width=0.99\hsize,height=0.35\textwidth,angle=0,clip]{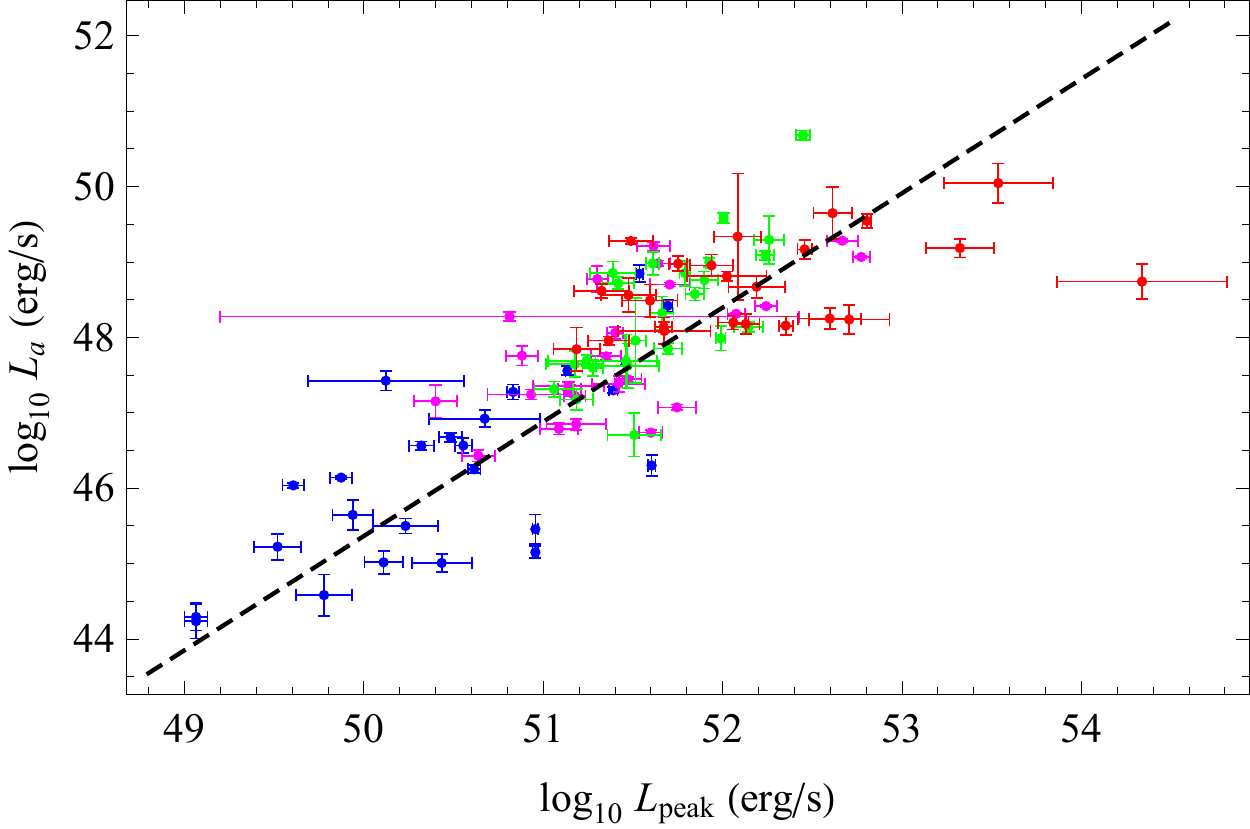}
\caption{GRB distributions in redshift bins at $L_a$--$L_{peak}$
 plane, where $L_{peak}$ is computed using the approach used in the Second BAT Catalog. The
sample is split-ed into 4 different equi-populated redshift bins: $z \le 0.84$
(blue), $0.84 \leq z < 1.8$ (magenta), $1.8 \leq z < 2.9$ (green) and $z \geq
2.9$ (red). The dashed line is the fitting correlation line.  
\label{fig5}}
\end{figure}

\begin{figure}
\includegraphics[width=0.99\hsize,height=0.30\textwidth,angle=0,clip]{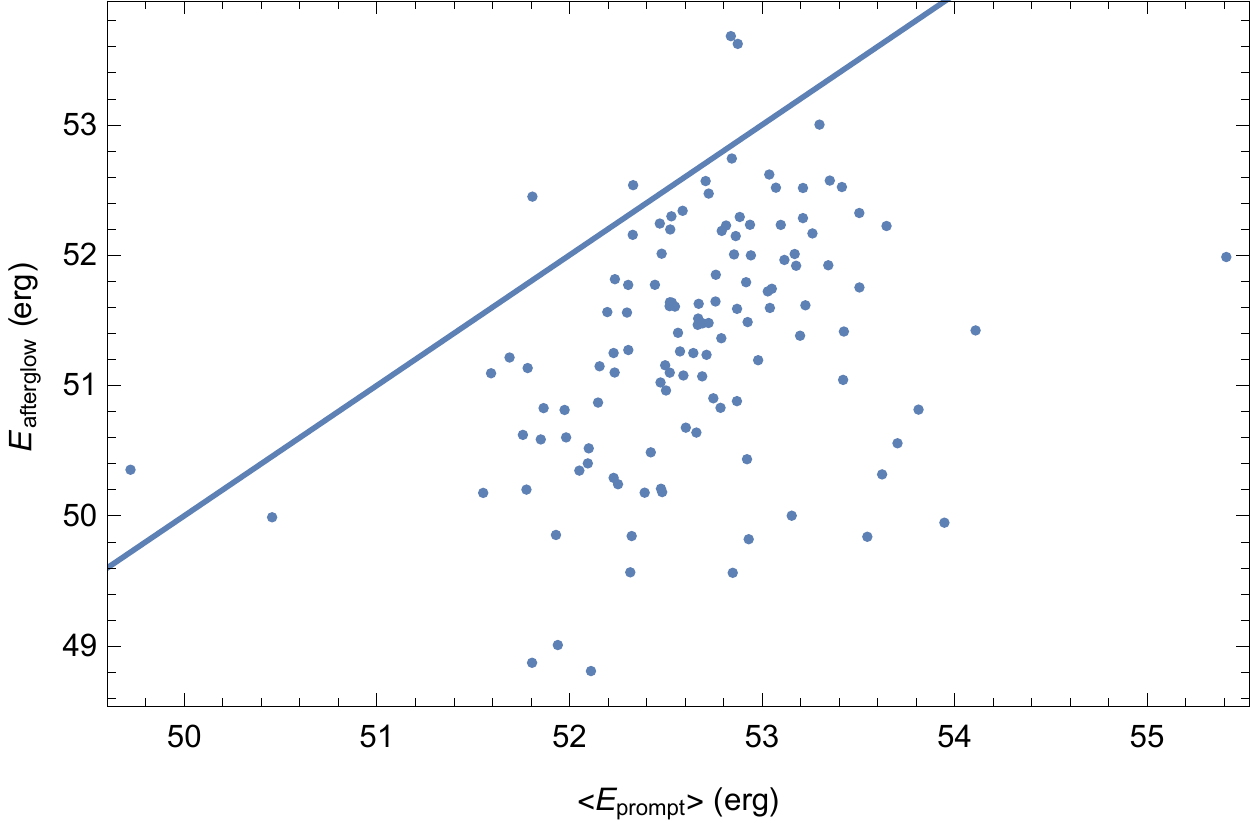}
\caption{Prompt averaged energy $<E_{prompt}>$ vs. afterglow energy,
$E_{afterglow}$, for 123 GRBs computed using the W010 model. The solid line for
equal prompt and afterglow energies is provided for reference.
\label{fig6}}
\end{figure}

{\bf At Fig. \ref{fig6} we compare the average prompt and the afterglow energies. The $ \langle E_{prompt} \rangle =\sum_{i=1}^{N} E_{pulse,i}/N$, where $E_{pulse,i}$ is the energy of each single pulse computed following Equ. \ref{eq: energy} in each GRB, N is the number of pulses in each GRB. For the afterglow the average afterglow energy, $<E_{afterglow}>$, coincides with $E_{afterglow}$ of the single pulses since we do not have multiple pulses in the afterglow in this sample, infact $N=1$ for each GRB afterglow.}
 Previously W07 found that in few cases $E_{afterglow} \equiv \langle E_{prompt} \rangle$, but in most cases $E_{afterglow}$ was roughly $10\%$ of the prompt emission. Here, with many more GRBs analyzed and within the pulse-afterglow model we confirm this result.  

The correlation of the prompt peak pulse isotropic luminosity averaged over all
single GRB pulses and the afterglow luminosity computed within the W010 model is
comparable with the one presented in the upper panel of Fig. \ref{fig5}, that
correlates $L_{peak}$, the isotropic peak luminosity of the brightest GRB
prompt emission pulse from the time of the burst, and $L_a$ where $L_{peak}$ has been computed using the
approach adopted in the Second BAT Catalog \citep{Sakamoto2011}, as described in \S 4.
We have tested over all the GRB
sample that $L_{peak}$, presented in Fig. \ref{fig5} (upper panel), has a consistent distribution compared to $L_{f}$, obtained from the
pulse fitting.

In Fig. \ref{fig5} we show that the correlation between $L_{peak}$ and $L_a$
exists even for different redshift bins.
The fitted correlation reads as follows:
\begin{equation}
\log L_{a}=A+B*\log L_{peak}
\label{Lpeak-La}
\end{equation}

 where $A=-14.67 \pm 3.46$ and $B=1.21^{+0.14}_{-0.13}$. 

Dainotti et al. (2011b) demonstrated that correlations exist between $L_a$ and
the luminosities for the prompt emission, computed as $E/T^{*}$, where $T^{*}$
are the characteristic GRB rest frame time scales $T^{*}_p=T_p/(1+z)$,
$T^{*}_{90}=T_{90}/(1+z)$ and $T^{*}_{45}=T_{45}/(1+z)$ \footnote{$T^*_{90}$ and
$T^*_{45}$ are the rest frame time scales for GRB energy emission between 5 and
95 \% and 5 and 50\% ranges of the total prompt emission respectively, while
$T^*_p$ is the rest frame time at the end of the prompt emission in the W07
model.}. 
We stress here that $\rho=0.79$ for the $L_{peak}-L_a$ correlation, where
$L_{peak}$ is computed according to the Second Bat Catalog, is considerably
increased compared to $\rho=0.60$ for the $L_{90}=E/T_{90}$ vs $ L_a$
correlation \citep{Dainotti2011b}. This means that a more suitable choice of the
parameters in the luminosities or energies definition can increase of the $24\%$
the correlation coefficient. We also note that here the sample is doubled
compared to the analysis performed by Dainotti et al. (2011b) in which the GRBs
analyzed were $62$.
In Fig. \ref{fig5} we selected the value of $L_{peak}$ computed from Eq.
\ref{eq: lx} assuming a broken power law or a simple power law as a spectral
model (as it has been explained in section \ref{data analysis}) thus not
involving error propagation due to time and energy as in the previous defined
luminosities. This is the reason why for this correlation we obtain an increment
of $\rho$.
 
We here underline the importance of the choice of the $L_{peak}$-$L_a$
correlation and not of the $E$-$L_a$ correlations presented in Dainotti et al.
(2011b), because $E$ may suffer from the systematic bias in duration
measurements. This would mean that although $E$ evolution studies may in fact be
biased at high redshift where a fraction of detected bursts grows with a low
signal-to-noise ratio, no such bias should exist for $L_{peak}$
\citep{Lloyd1999}. Therefore, the luminosity-duration is more reliable than the
energy-duration correlation, and in the present paper this is the reason why we
addressed the attention to the $L_{peak}$-$L_a$ relation, instead of $E-L_a$.

\section{The redshift dependence}

\begin{figure}
\includegraphics[width=0.99\hsize,height=0.35\textwidth,angle=0,clip]{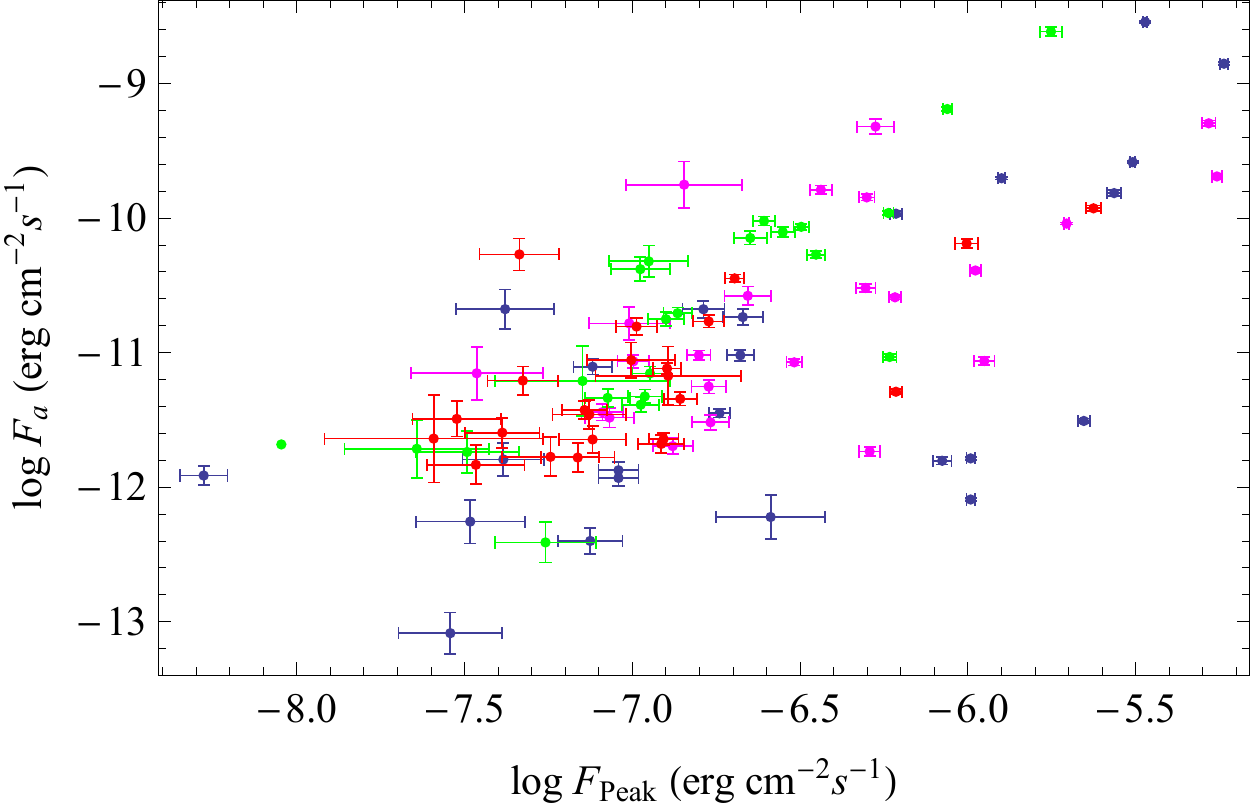}
\caption{GRB distributions in redshift bins at the ${\rm F_a}$--$\rm F_{peak}$ plane, where $F_{peak}$ 
is computed following the approach used in the Second BAT Catalog. The
sample is split-ed into 4 different equi-populated redshift bins: $z \le 0.84$
(blue), $0.84 \leq z < 1.8$ (magenta), $1.8 \leq z < 2.9$ (green) and $z \geq
2.9$ (red). The dashed line is the fitting correlation line. 
\label{fig7}}
\end{figure} 

The $L_{peak}-L_a$ correlation could be due to the dependence of luminosity on distance, since
it involves two luminosities. We compare Fig. \ref{fig5} and Fig. \ref{fig7}
in order to clarify how much this dependence influences the existence
of the correlation itself. In support of the existence of the $L_{peak}$-$L_a$
correlation we show the correlation between observed fluxes $F_a$, the flux at
time $T_a$, vs. the peak flux in the prompt emission, $F_{peak}$,
$F_a$-$F_{peak}$, with a Spearman correlation coefficient $\rho=0.63$ (see
 Fig. \ref{fig7}). Thus, we remove with a first rough approximation the redshift dependence induced
by the distance luminosity using fluxes instead of luminosities. In fact,
if the $L_{peak}-L_a$ correlation was completely due to the induced redshift dependence
this would have caused a disappearing of the correlation or a drastically
reduced value of $\rho$ less than $0.5$ and a probability of occurrence by
chance $>5\%$, which is not the case.
Then, to evaluate the presence of redshift evolution we follow the approach
adopted in Dainotti et al. (2011a, 2013a) by dividing the sample into $4$ redshift
bins. The GRBs distribution in each redshift bin is not clustered or confined in
a given subspace, see Fig. \ref{fig5}, thus suggesting no strong redshift
evolution. This is expected for $L_a$, because Dainotti et al. (2013a)
demonstrated that there is no redshift evolution of this luminosity. However,
Petrosian et al. (2015) show that $L_{peak}$ is affected by the redshift
evolution as $L_{peak}/(1+z)^{2.3}$ using a more complex function than the simple power law, used previously for GRBs \citep{Dainotti2013a}. 
Here the sample has been chosen differently from Petrosian et al. (2015), because only observations which 
have good coverage of the data in the early prompt and can be fitted within the W010 model are taken into account.
Therefore, for a more precise evaluation we have to address the problem of the luminosity evolution for this specific sample.  
For a quantitative analysis of this problem we apply the Efron and Petrosian (1992) method. 

\section{The Efron and Petrosian method}

\begin{figure}
\includegraphics[width=1.0\hsize,angle=0,clip]{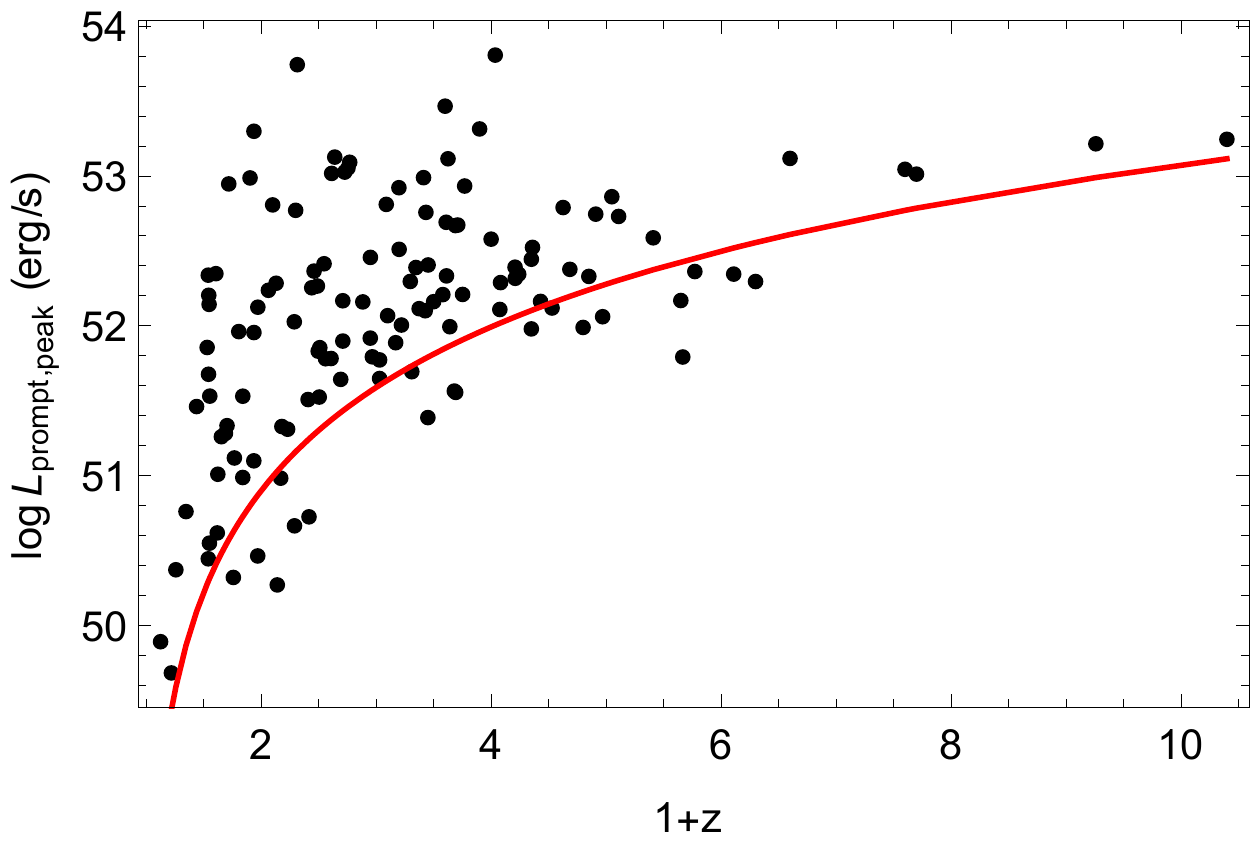}
\includegraphics[width=1.0\hsize,angle=0,clip]{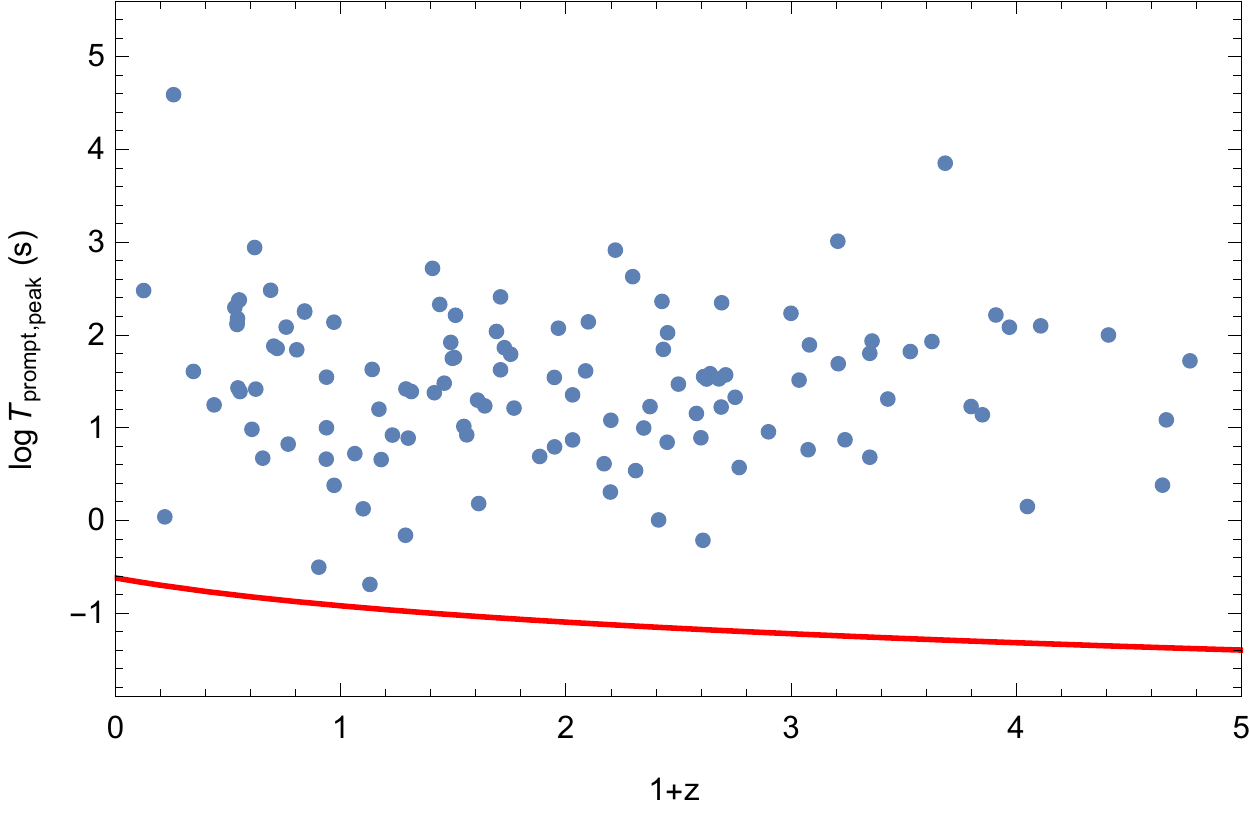}
\caption{Upper Panel: The bivariate distribution of $L_{peak}$ and redshift with the flux limit assuming the K correction $K=1$. The BAT flux limit, $4.0 \times 10^{-8}$ erg cm$^{-2}$ (solid red line) which better represents the limit of the sample. 
{\bf Lower panel}: The bivariate distribution of the rest frame time $T^*_{prompt}$ and the redshift, where here with $T_{prompt}$ we denote the sum of the peak pulses width of each single pulse in each GRB. The chosen limiting value of the observed pulse width in the sample, $T_{prompt,lim}= 0.24$ s. The red line is the limiting rest frame time, $T_{prompt,{\rm lim}}/(1+z)$.}
\label{fig8}
\end{figure} 

The first important step for determining the distribution of true correlations among
the variables is the quantification of the biases introduced by the observational selection effects due to the selected sample and the instrumental limits.
In the case under study the selection effect or bias that distorts the statistical correlations are the flux limit and the temporal resolution of the instrument.
To account for these effects we apply the Efron \& Petrosian technique, already successfully applied for GRBs \citep{Petrosian2009,Lloyd2000,Kocevski2006}. 
The EP method reveals the intrinsic correlation because the method is specifically designed to overcome the biases resulting from incomplete data. Moreover, it identifies and removes also the redshift evolution present in both variables, time and luminosity.

The EP method uses a modified version of the Kendall $\tau$ statistic to test the independence of variables in a truncated data.
Instead of calculating the ranks $R_{i}$ of each data points among all observed objects, which is normally done for an untruncated data, the rank of each data point is determined among its ``associated sets" which include all objects that could have been observed given the observational limits. 

Here we give a brief summary of the algebra involved in the EP method. This method uses the Kendall rank test to determine the best-fit values of parameters describing the correlation functions using  the test statistic 

\begin{equation}
\tau = {{\sum_{i}{(\mathcal{R}_i-\mathcal{E}_i)}} \over {\sqrt{\sum_i{\mathcal{V}_i}}}}
\label{tauen}
\end{equation}
to determine the independence of two variables in a data set, say ($x_i,y_i$) for  $i=1, \dots, n$.  Here $R_i$ is the rank of variable $y$ of the data point $i$ in a set associated with it. For a untruncated data (i.e. data truncated parallel to the axes) the {\it associated set} of point $i$ includes all of the data with  $x_j < x_i$.  If the data is truncated one must form the {\it associated set} consisting only of those points which satisfy conditions imposed by the limiting instrumental values, see definition below. 

If ($x_i,y_i$) were independent then the rank $\mathcal{R}_i$ should be distributed continuously between 0 and 1 with the expectation value $\mathcal{E}_i=(1/2)(i+1)$ and variance  $\mathcal{V}_i=(1/12)(i^{2}-1)$. Independence is rejected at the $n \, \sigma$ level if $\vert \, \tau \, \vert > n$.  
Here the mean and variance are calculated separately for each
associated set and summed accordingly to produce a single value for
$\tau$. This parameter represents the degree of correlation for the
entire sample with proper accounting for the data truncation.

With this statistic, we find the parametrization that best describes the luminosity and time evolution for the prompt emission. For the afterglow emission we refer to results already presented in Dainotti et al. (2013a).
We now have to determine the limiting flux, $F_{lim}$, which gives the minimum observed luminosity for a given redshift, $L_{lim}=  4 \pi D_L^2(z) \, F_{lim} K$. At the upper panel of Fig. \ref{fig8} we show the limiting luminosity for $K=1$ just not to show fuzzy boundaries, but for an appropriate evaluation of the luminosity evolution we assign to each GRB its own K correction. We have investigated several limiting fluxes to determine a good representative value, while keeping an adequate size of the sample itself. We have finally chosen the limiting flux $F_{lim} = 4.0 \times $10$^{-8}$ erg cm$^{-2}$, which allows $116$ GRBs in the sample. We have also chosen the observed minimum pulse width of the prompt, which is $T^{*}_{prompt,{\rm lim}}= 0.24/(1+z)$ s, lower panel of Fig. \ref{fig8}. This time has been computed as the sum of the single pulses width in each GRB. In such a way we can employ a comparison with previous time evolution in the afterglow as presented in \cite{Dainotti2013a}.

\subsection{The luminosity and time evolutions}

For the luminosity and time evolution it is necessary to first determine whether
the variables $L_{peak}$ and $T^*_{prompt}$, are correlated with redshift or are
statistically independent. For example, the correlation between $L_{peak}$ and the redshift, $z$, is what we
call luminosity evolution, and independence of these variables
would imply absence of such evolution. 
The EP method prescribed how to remove the correlation by defining
new and independent variables.

We determine the correlation functions, $g(z)$ and $f(z)$ when determining the evolution of $L_{peak}$ and $T^{*}_{prompt}$ so that de-evolved variables, namely the local variables, $L'_{peak} \equiv L_{peak}/g(z)$ and $T'_{prompt} \equiv T^*_{prompt}/f(z)$ are not correlated with z. 
The evolutionary functions are parametrized both by simple correlation functions or more complex ones.

The simple power law functions are represented by 

\begin{equation}
g(z)=(1+z)^{k_{L_{peak}}}, f(z)=(1+z)^{k_{T^{*},prompt}}
\label{lxev}
\end{equation}
 
so that $L'_{peak}=L_{peak}/g(z)$ refer to the local ($z=0$) luminosities. The more complex function chooses a fiducial critical Z, where we define $Z=1+z$. We chose $Z_{cr}=3.5$, thus allowing the following functional form for

\begin{equation}
 g(z)=\frac{Z^{k_{L}}(1+Z_{cr}^{k_L})}{Z^{k_L}+Z_{cr}^{k_L}}, f(z)=\frac{Z^{k_T^{*}}(1+Z_{cr}^{k_T^{*}})}{Z^{k_T^{*}}+Z^{k_T^{*}}_{cr}}
\label{lxev2}
\end{equation}

We computed both approaches obtaining compatible results.
The associated set for the source $i$ to obtain the luminosity evolution is :
\begin{equation} \label{eq:Ji1}
    J_{i} \equiv \{j:L_{j} > L_{min}(i) \}  \vee  \{j:   L_j > L_i \}  \vee  \{j: z_j < z_i\},
   \end{equation}
 
where $L_{min}(i)$ is the minimum luminosity of the object $i$ corrispodent to $L_i$, $z_i$ is the redshift of the object $i$.
The objects of all the sample are indicated with $i$, while the objects in the associated sets are denoted with $j$.  With the the simbol $\vee$ we indicate the union of the sets.

\begin{figure}
\includegraphics[width=0.50\textwidth,height=0.48\textwidth]{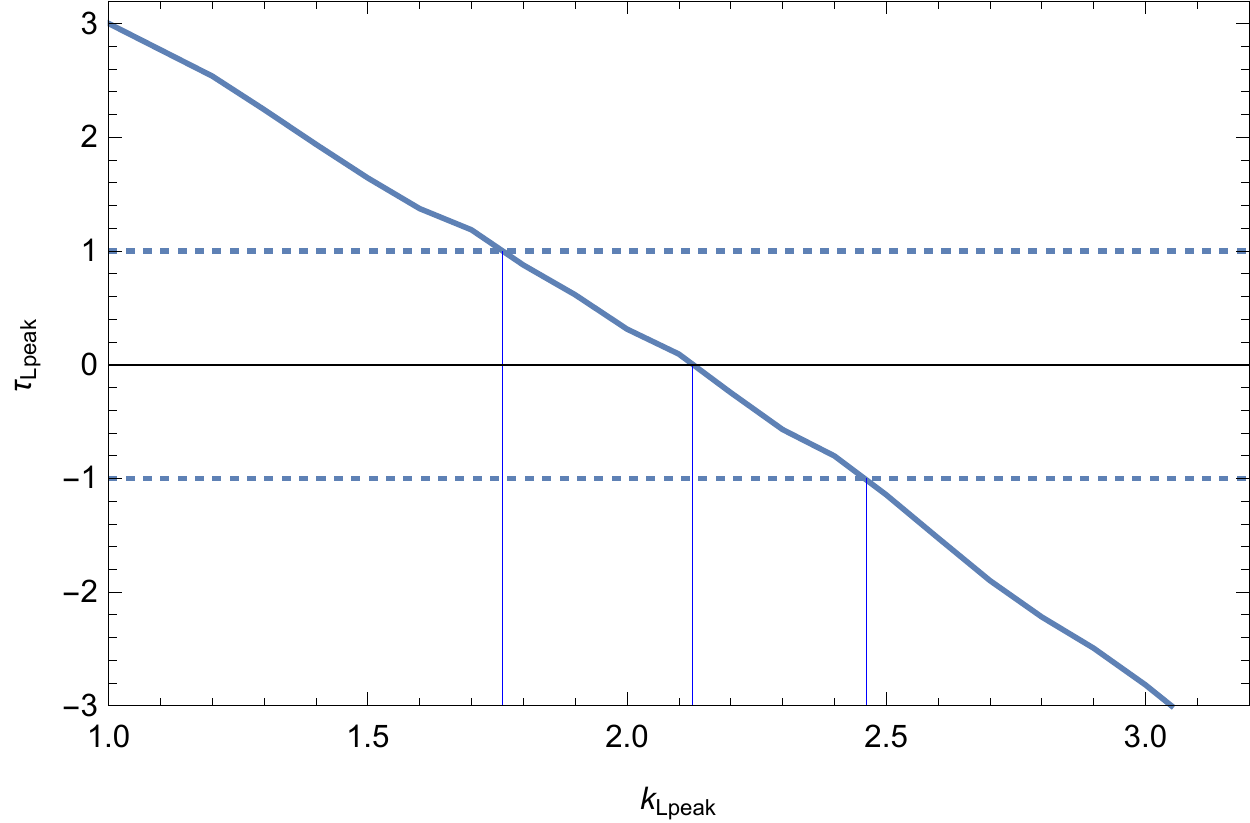}
\includegraphics[width=0.50\textwidth,height=0.48\textwidth]{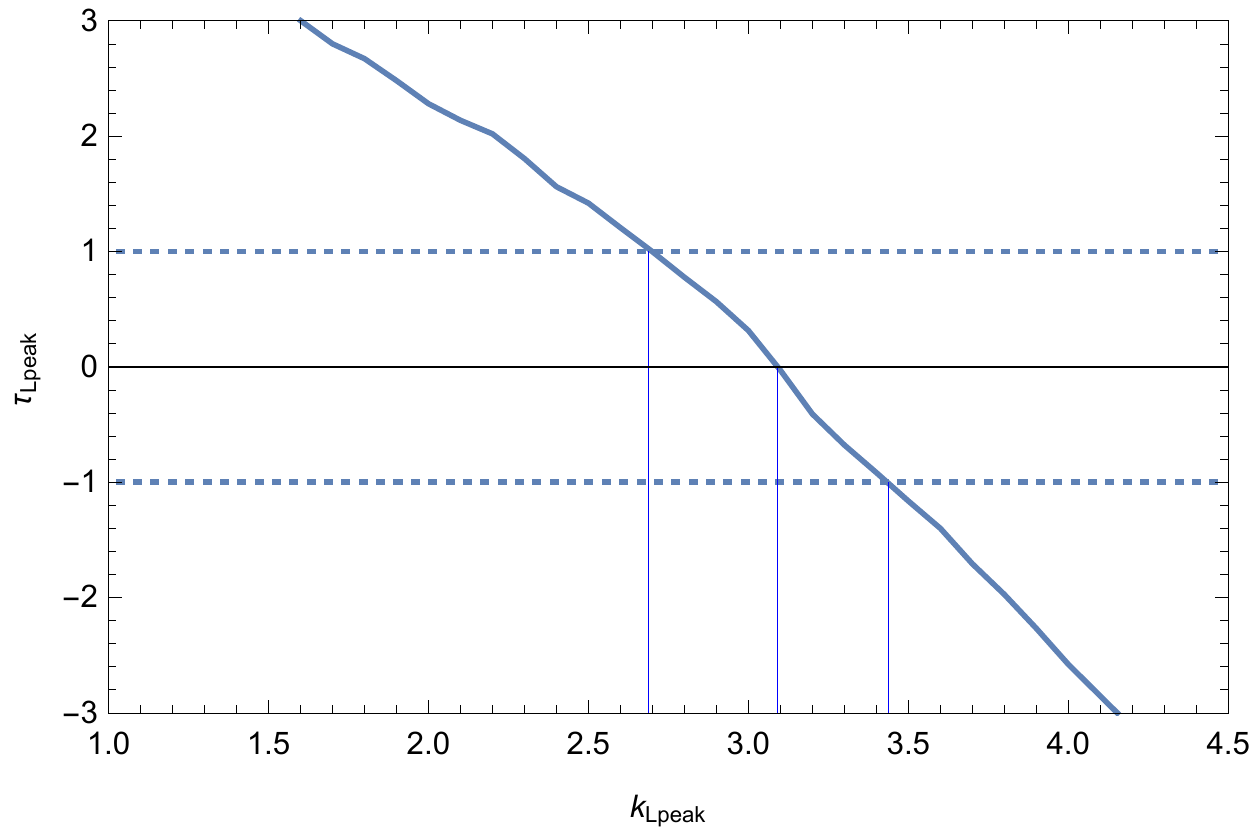}
\caption{ {\bf Upper:} Test statistic $\tau$ vs. $k_{L_{peak,prompt}}$, the luminosity evolution defined by Eq. \ref{lxev} using a simple power law as $g(z)$.{\bf Lower:} The same test statistic using a more complex function for the evolution $g(z)$, defined by the Eq. \ref{lxev2}.}
 \label{Fig9}
\end{figure}

Analogously, to obtain the pulse width evolution factor we need to compute the associated set for a given object $i$, which are :

\begin{equation} \label{eq:Ji2}
    J_{i} \equiv \{j:  T_j > T_{min,i} \}  \vee  \{j: T_j > T_i  \} \vee \{j: z_j > z_i \},
\end{equation}

\noindent where $T_{\rm min}(T_{prompt,i})$ is the minimum $T_{prompt}$ at which object $i$ could be still included in the survey given its peak width duration and the limiting time of the observation.

\begin{figure}
\includegraphics[width=0.50\textwidth,height=0.50\textwidth]{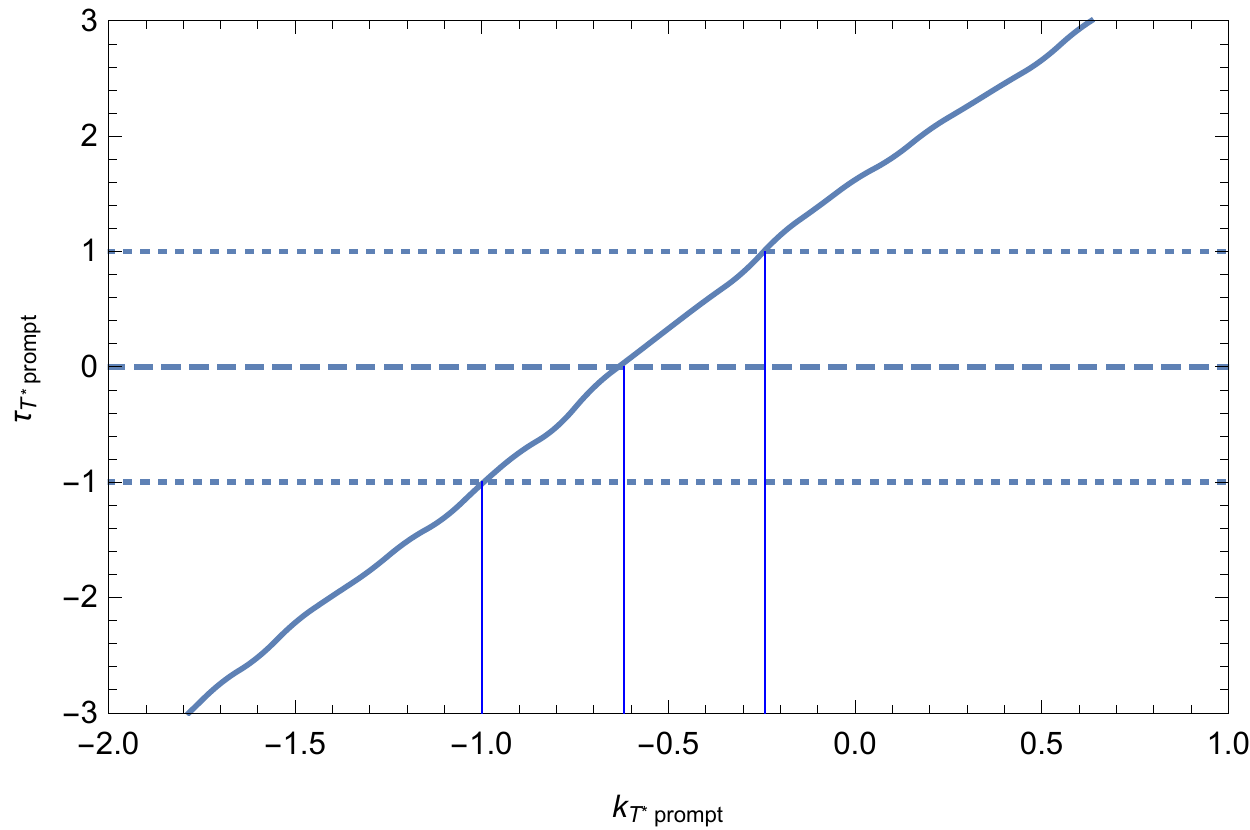}
\includegraphics[width=0.50\textwidth,height=0.50\textwidth]{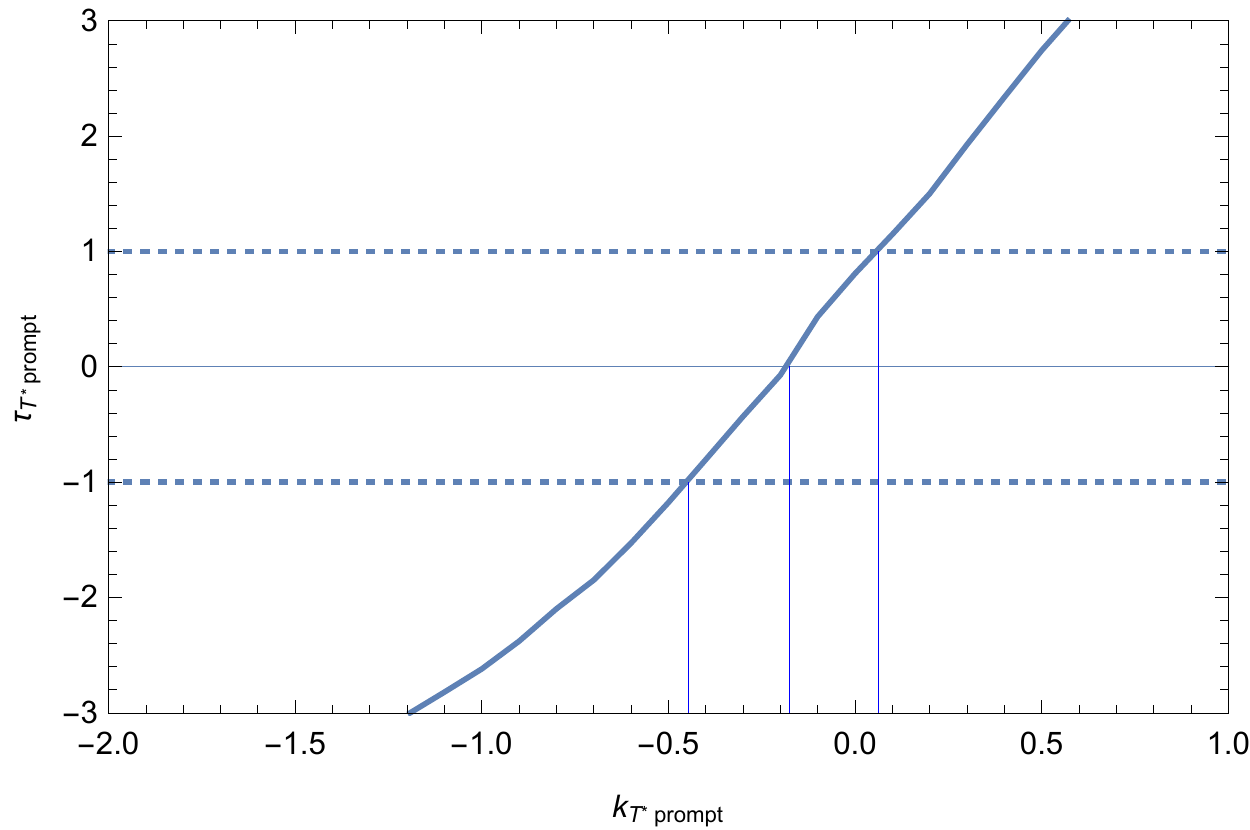}
\caption{Upper panel: Test statistic $\tau$ vs. $k_{T^{*}_{prompt}}$, the time evolution defined by Eq. \ref{lxev}, Lower panel: The same test statistic using a more complex function for the evolution $g(z)$, defined by the Eq. \ref{lxev2}}
\label{Fig10}
\end{figure}

With the specialized version of Kendell's $\tau$ statistic, the values of $k_{L_{peak}}$ and $k_{T^{*}_{prompt}}$ for which $\tau_{L_{peak}} = 0$ and $\tau_{T^{*}prompt} = 0$ are the ones that best fit the luminosity and width pulse evolution respectively, with the 1$\sigma$ range of uncertainty given by $| \tau_x | \leq 1$. Plots of $\tau_{L_{peak}}$ and $\tau_{T^{*}_{prompt}}$ versus $k_{L_{peak}}$ and $\tau_{T^{*}_{prompt}}$ are shown in Fig. \ref{Fig9} and Fig. \ref{Fig10} respectively. With $k_{L_{peak}}$ and $k_{T^{*}prompt}$ we are able to determine the de-evolved observables $T{'}_{prompt}$ and $L{'}_{peak}$. 

There is a significant luminosity evolution in the prompt, $k_{L_{peak}}=2.13_{-0.37}^{+0.33}$, and much less significant in the time, $k_{T^{*}_{prompt}}=-0.62 \pm 0.38$ for the simple power law functions. If we consider the more complex function for the evolution we obtain $k_{L_{peak}}=3.09_{-0.35}^{+0.40}$ and $k_{T^{*}_{prompt}}=-0.17_{-0.27}^{+0.24}$.
It is straightforward that we achieve an higher evolution for luminosity and a smaller evolution for the time for the way we chose the function.
We also note that the results of the luminosity evolutions among the two different functions are compatible within $2 \sigma$, while the time evolutions are compatible within 1 $\sigma$.

\subsection{The intrinsic $L_{peak}-L_a$ correlation}
\begin{figure}
\includegraphics[width=0.50\textwidth,height=0.50\textwidth]{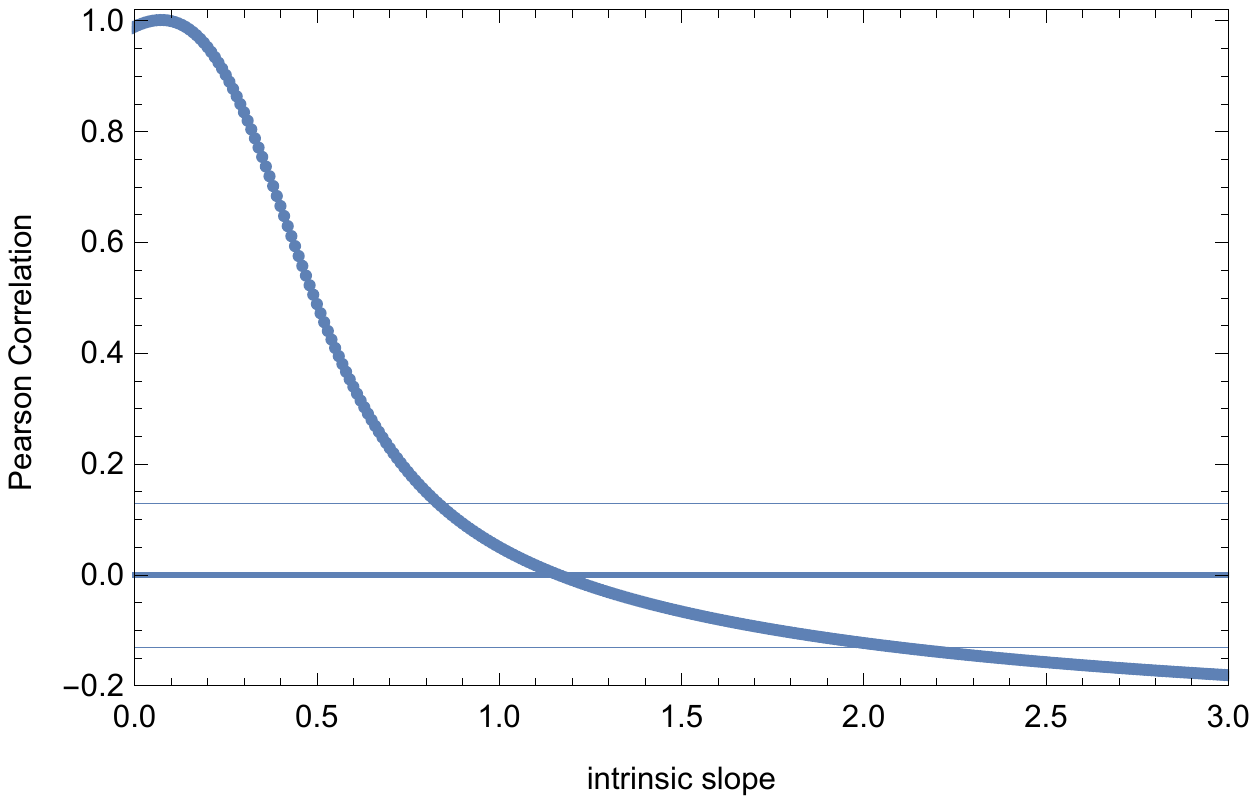}
\caption{Local luminosity-luminosity correlation coefficient vs the intrinsic slope showing the best value where $L^{'}_{peak}$ and $L^{'}_a$ are significantly correlated (the central thick line). The two thinner lines parallel to $r=0$ shows the $0.05\%$ probability that the sample is drawn by chance.}
\label{Fig11}
\end{figure}

\footnote{Here we do not consider the de-evolved $L^{'}_{peak}-T^{'}_{prompt}$ correlation because the $T^{'}_{prompt}$ adopted is the sum of the all time widths of all the pulses for each GRB and not the width of the single pulse. Therefore, we cannot determine with accuracy the evolution in time for the prompt since for single pulses we are not able to apply the Efron and Petrosian method, because we have only $1$ limiting time for all the total integrated time over all the pulses and this does not coincide with the minimum time among each single pulse. Thus, this discrepancy in the limiting time determination can lead to an inaccuracy in the evaluation of the time evolution.
Notwithstanding this difficulty for the time evolution, for the luminosity evolution this problem does not occur, since we chose the maximum peak luminosity of each GRB among the all pulses in that given GRB.}
We here focus on determining the intrinsic correlation among the local luminosities $L^{'}_{peak}-L^{*}_a$. Following the method presented in Petrosian \& Singal (2014) we compute the dependence of this correlation from the luminosity distance.
According to Eq. \ref{Lpeak-La} we can rename the variables with an abuse of notation for simplicity as $\log L^{'}_a=L^{'}_a$, $\log L^{'}_{peak}=L^{'}_{peak}$ and $\log D_L=D_L$ in order to write in a simpler way the partial correlation coefficient in the log space domain:

\begin{equation}
r_{L^{'}_{peak} L^{'}_a, D_L}=\frac{r_{L^{'}_{peak},L^{'}_a}-r_{L^{'}_{peak},D_L}*r_{L^{'}_a,D_L}}{(1-r^{2}_{L^{'}_{peak},D_L})*{(1-r^{2}_{L^{'}_a,D_L})}}
\end{equation}

which accounts for mutual distance dependence of the luminosities. We now consider the correlation in the local luminosity space so that $L^{''}_{a}=L^{'}_{peak} - \alpha L^{'}_a$ and we calculate the $r_{L^{'}_{peak}, L^{'}_a, D_L}$ as a function of the index $\alpha$, namely the intrinsic slope. As shown in Fig. \ref{Fig11} the correlation becomes significant for $\alpha=1.14^{0.83}_{-0.32}$, which is very close to the observed correlation. The errorbars quoted are at the $2$ $\sigma$ significance level.

\section{Summary and Discussion}

The analysis presented in this study reveals that

\begin{itemize}
\item{prompt and plateau phases dissipate similar amounts of energy, but over very different time
scales as shown through the figures \ref{fig1}, \ref{fig2} and \ref{fig6}}. 
\item{slopes in the luminosity-duration distributions between the prompt and
plateau emissions $L_{f}-T^{*}_{f}$ vs $L_a-T^{*}_a$ differ almost $3$
$\sigma$, while in the local luminosity space more than 3 $\sigma$. However, for the evaluation of
the time evolutions of the pulse in the prompt there is the problem of determining the proper 
limiting time of the pulses, as we explained in footnote 8. Therefore, a definite conclusion on the differences in the slopes 
still needs to be reached and this will be object of a forthcoming investigation.
 The evidence of difference between prompt and afterglow is then recalled also by the difference in the spectral
parameters of the prompt and the afterglow phases. Also this fact does not imply necessarily a diverse
mechanism as we have pointed out in \S 5.2.} 
\item{The extended luminosity-duration distributions $L_{f}-T^{*}_{f}$, see upper
panel of Fig. \ref{fig2} and the energy-duration correlation, see the middle
panel of Fig. \ref{fig2} show that there is continuity in transition from prompt
distribution to the afterglow one, namely no gap in the data. Difference between
the prompt and plateau slopes is present independently from the choice of
luminosity or energy. The luminosity-duration and energy-duration spaces are just
two ways of looking at the same data, as well as the difference in the
correlations. The $E_{total}$-duration plot in the lower panel of Fig. \ref{fig2}
clearly shows that the plateaus occupy a different area of the energy-duration
plane to the pulses.
Individual prompt pulses and plateaus both produce energy values in the same
broad range, but the plateau duration is on average a factor of $100$ larger.}
\item{Stronger correlations are present when we compare respectively $<L_{prompt}>-L_{a}$ and
$L_{peak}-L_a$ luminosities, see Fig. \ref{fig5}, rather than considering $L_a$ and the prompt
emission luminosities computed as ratio of energy over a particular time scale,
such as $L_{45}=E/T_{45}$ and $L_{90}=E/T_{90}$, \citep{Dainotti2011b}.}

\item{We found very interestingly that the $L_{peak}-L_a$ correlation is very robust also in the local luminosity space when 
we removed the luminosity evolution both in the prompt and in the afterglow 
and it presents a compatible result of the intrinsic slope with the observed slope within 1 $\sigma$. 
This will have impact on the investigation for the theoretical models.}

\end{itemize}

From this analysis we hypothesize that

\begin{itemize}
\item{Both the different slopes in the luminosity-duration and in the energy-duration
space of prompt pulses and plateau ones might indicate that these two are quite
distinct features of the emission. The former probably come from internal shocks and
the latter from the external shock. The prompt pulses are fast cooling while the plateau pulses are slow
cooling. This is known from the literature for the prompt and afterglow phases,
\citep{RM94,RM98}, but the upper panel of Fig. \ref{fig2} shows that this
statement might be true also for the plateau phase. So this is another significant difference
between the prompt and plateau phase indicating that if the latter is due to
synchrotron from the external shock (which is likely) then the pulses all have
very similar physical conditions in the shock. In particular, the power law index
of the electron distribution is very similar in all cases.}
\item{The present study is relevant to quantify the mentioned
relations in order to improve or modify the
existing physical model of GRB emission which should predict 
the $L_{peak}$ vs. $L_a$ correlation together with the combined luminosity-time
correlations both in prompt and afterglow phases. In particular, among the models we have mentioned in the theoretical motivation of this work the one that better describe the observed correlations is the model by Hascoet et al. (2014), because some particular configurations of the microphysical parameters are able to reproduce the luminosity-time correlations difference in slopes and the $<L_{prompt}>-L_{a}$ correlations. Also the model proposed by Ruffini et al. (2014) is able to reproduce these observational features, while thin shell models, \citep{Erten2014a}, are ruled out.}
 
\end{itemize}
In conclusion, all these observational evidences taken into
account contemporaneously are able to better test and discriminate some of the existing theoretical models.

\section{Acknowledgments}

This work made use of data supplied by the UK Swift Science Data Centre at the
University of Leicester. We are grateful to M. Barkov and F. Rubio Da Costa for useful comments and
remarks on the present manuscript. M.G.D. is grateful for the initial support
from the JSPS (No. 25.03786). Moreover, the research leading to these result has received funding from the European Union Seventh Framework Programme (FP7/2007-2013) under grant agreement n 626267. M.G.D. and S.N. are grateful to the iTHES Group discussions at Riken. M.G.D is also grateful to
R. W. and P. O. to be hosted by the Astronomy Department at Leicester University
through their Department grant. M.O. is grateful to the Polish National Science
Centre for support through the grant DEC-2012/04/A/ST9/00083. S. N. is grateful
to JSPS (No.24.02022, No.25.03018, No.25610056,  No.26287056) $\&$
MEXT(No.26105521).

\appendix
\section{The D'Agostini fitting method}
We briefly present the D{'\ Agostini method \citep{Dago05}, used to fit the above
mentioned correlations. This takes into account the intrinsic scatter, thus
providing more reliable errors. 
Let us suppose that $R$ and $Q$ are two quantities related by a linear relation

\begin{equation}
R = a  Q + b
\end{equation}
and denote with $\sigma_{int}$ the intrinsic scatter around this relation.
Calibrating such a relation means determining the two coefficients $(a, b)$ and
the intrinsic scatter $\sigma_{int}$. To this aim, we will resort to a Bayesian
motivated technique \cite{Dago05} thus maximizing the likelihood function
${\cal{L}}(a, b, \sigma_{int}) = \exp{[-L(a, b, \sigma_{int})]}$ with\,:

\begin{equation}
L(a, b, \sigma_{int}) = \frac{1}{2} \sum{\ln{L_1}} + \frac{1}{2} \sum{\ln{L_2}}
\label{eq: deflike}
\end{equation}

where
\begin{equation}
L_1=(\sigma_{int}^2 + \sigma_{R_i}^2 + a^2 \sigma_{Q_i}^2) 
\end{equation}
and 
\begin{equation}
L_2=\frac{(R_i - a Q_i - b)^2}{\sigma_{int}^2 + \sigma_{Q_i}^2 + a^2
\sigma_{Q_i}^2}
\end{equation}

where the sum is over the ${\cal{N}}$ objects in the sample.
The above formulae easily applies to our case setting $R = \log L^*_X(T_a)$ and
$Q = \log T^*_a$. We estimate the uncertainty on $\log L^*_X(T_a)$ by
propagating the errors on $(T_a, F_a, \beta_a)$.

The Bayesian approach used here also allows us to quantify the uncertainties on
the fit parameters. To this aim, for a given parameter $p_i$, we first compute
the marginalized likelihood ${\cal{L}}_i(p_i)$ by integrating over the other
parameter. The median value for the parameter
$p_i$ is then found by solving\,:

\begin{equation}
\int_{p_{i,min}}^{p_{i,med}}{{\cal{L}}_i(p_i) dp_i} = \frac{1}{2}
\int_{p_{i,min}}^{p_{i,max}}{{\cal{L}}_i(p_i) dp_i} \ .
\label{eq: defmaxlike}
\end{equation}
The $68\%$ ($95\%$) confidence range $(p_{i,l}, p_{i,h})$ are then found by
solving\,:

\begin{equation}
\int_{p_{i,l}}^{p_{i,med}}{{\cal{L}}_i(p_i) dp_i} = \frac{1 - \varepsilon}{2}
\int_{p_{i,min}}^{p_{i,max}}{{\cal{L}}_i(p_i) dp_i} \ ,
\label{eq: defpil}
\end{equation}

\begin{equation}
\int_{p_{i,med}}^{p_{i,h}}{{\cal{L}}_i(p_i) dp_i} = \frac{1 - \varepsilon}{2}
\int_{p_{i,min}}^{p_{i,max}}{{\cal{L}}_i(p_i) dp_i} \ ,
\label{eq: defpih}
\end{equation}
with $\varepsilon = 0.68$ (0.95) for the $68\%$ ($95\%$) range respectively.

The $a$ and $b$ parameters are independent and the computation of the error is
performed around the actual variable and not in the barycenter of points. 

\end{document}